\documentstyle[10pt,emulateapj]{article}
\def\deg{$^{\circ}\,$}
\def\solm{M$_{\odot}\,$}
\def\soll{L$_{\odot}\,$}
\def\kms{km s$^{-1}$}

\begin{document}

\title{Galaxy Populations and Evolution in Clusters III.  The Origin of
Low-Mass Galaxies in Clusters: Constraints from Stellar Populations}

\author{Christopher J. Conselice$^{1}$,  John S. Gallagher, III$^{2}$, 
Rosemary F.G. Wyse$^{3}$}

\altaffiltext{1}{California Institute of Technology, Mail Code 105-24, Pasadena, CA}
 
\altaffiltext{2}{Department of Astronomy, University of Wisconsin, Madison 
475 N. Charter St. Madison, WI, 53706-1582}


\altaffiltext{3}{Department of Physics \& Astronomy, The Johns Hopkins University}

\begin{center}
{\it Accepted to the Astronomical Journal}
\end{center}

\begin{abstract}

Low-mass galaxies in nearby clusters are the most numerous galaxy type in
the universe, yet their origin and properties remain largely unknown.  To
study basic questions concerning these galaxies we present the results of a 
survey designed to constrain the characteristics and properties of the 
stellar populations in a magnitude complete sample of low-mass cluster 
galaxies (LMCGs) in the center of the Perseus cluster.   Using 
deep, high-quality WIYN UBR images to obtain photometric and structural 
properties, we demonstrate that the
53 LMCGs in our sample have a significant scatter about
the color-magnitude relationship at M$_{\rm B} > -15$.
By comparing single-burst stellar population models to
our photometry, we argue that
stellar populations in LMCGs all have ages $>$ 1 Gyrs,
with redder LMCGs containing stellar metallicities 
[Fe/H] $> -0.5$.  By assuming that the colors of LMCGs reflect metallicity,
and have co-evolved with the giant ellipticals, we find a wide range of 
values, from solar to [Fe/H] $\sim -3$.  We argue from this that
LMCGs have multiple origins, and fundamentally differ
from Local Group dwarf spheroidals/ellipticals.  The inferred 
lower metallicities of the bluer LMCGs implies that these are possibly 
primordial galaxies formed through 
self-enrichment and stellar feedback provided by winds from supernova.
We also investigate several other formation scenarios for these LMCGS, 
including: self-enrichment induced by the confinement of 
metals in halos by the intracluster medium, in situ formation out of
intracluster gas, systems with extreme dark halos, and as remnants of
previously higher mass systems.   We conclude that
roughly half of all low-mass cluster galaxies in the center of
Perseus have stellar populations and kinematic
properties, as discussed in previous papers in this series, consistent
with being the remnants of stripped galaxies accreted into clusters several 
Gyrs ago.

\end{abstract}

\section{Introduction}

Faint low-mass galaxies, especially those found in groups and
clusters, are the most common galaxy type in the local universe, yet
their nature remains elusive  (e.g., Ferguson \& Binggeli 1994).  
Because they are so common, these low-mass cluster galaxies
(LMCGs) likely hold answers to some of the ultimate
questions of how galaxies form and evolve.  They are also the least
massive galaxies known, whose nature is important for
understanding hierarchical galaxy formation 
(e.g.,  White \& Frenk 1991). 

In Cold Dark Matter (CDM) cosmological models, 
containing a bottom up formation scenario for structure (e.g. White
\& Rees 1978), low-mass galaxies should be the oldest objects in the
universe from which all other galaxies formed.  While observations
now reveal that stars in low-mass galaxies are potentially younger, on average,
than giant elliptical galaxies (Rakos et al. 2001) there are a plethora of 
theories, both cosmological
and evolutionary, for explaining how low-mass galaxies formed, generally
within the CDM milieu (e.g., Dekel
\& Silk 1986; Silk, Wyse \& Shields 1987; Babul \& Rees 1992; 
Kepner, Babul \& Spergel 1997; Mao \& Mo 1998; Moore
et al. 1998; Natarajan, Sigurdsson \& Silk 1998; 
Ferrera \& Tolstoy 2000; Cen 2001)

The traditional approach to studying low-mass galaxies is to examine
them in the Local Group (LG).  It is now well established that LG
dwarf elliptical and dwarf spheroidal galaxies  
have varying star formation
histories, with metal-poor populations as old as classical halo globulars, but
also with evidence for star formation occurring as
recently as 2-3 Gyrs ago (see Mateo 1998 and van den Bergh 2000 for reviews). 
There are also some low-mass LG galaxies, such as Sagittarius, that contain 
surprisingly metal-rich populations given their luminosities (e.g.,
Ibata, Gilmore \& Irwin 1995).   Because LG dEs are close
we can resolve 
their stellar populations, and thus we know much more about them than we do 
LMCGs, including their internal kinematics and star formation 
histories.  Many of the lowest mass galaxies in the LG, such as Draco and Ursa 
Minor, have very high inferred central $M_{\rm total}/L$ ratios (Kleyna et 
al. 2002), and apparently contain the densest dark matter 
halos of all known galaxies, in qualitative agreement with the original
CDM predictions (e.g., Lake 1990).  Low-mass galaxies
in clusters however have different kinematic and spatial properties than 
these LG systems (e.g.,
Conselice et al. 2001a) suggesting they might have a different formation
scenario.

We can learn more about the formation of low-mass galaxies and
their role in cosmology and large mass galaxy evolution by studying them  
where they are most abundantly found, in galaxy clusters.   
The high abundance of 
low-mass, or dwarf, galaxies in clusters was first
pointed out by Reeves (1956), who studied them in the Virgo cluster,
and by Hodge (1959, 1960) in the Fornax cluster.  Soon after it was
realized that both dwarf irregular-like (gas rich systems with star formation)
and dE-like (gas poor systems with little star formation) objects
exist in clusters (Hodge, Pyper \& Webb 1965).  Later, it became clear
that these dwarfs are not clustered around giant galaxies as they
are in groups (Reeves 1977), revealing perhaps a fundamental
difference in origin.  The modern study of LMCGs 
began with detailed optical and infrared observations of
individual Virgo dwarfs (e.g., Bothun \& Caldwell 1984; Bothun et al. 1985;
Ichikawa, Wakamatsu \& Okamura 1986; Gallagher \& Hunter 1986; Bothun
\& Mould 1988), and extensive catalogs of dwarfs in nearby clusters
(e.g., Bingelli, Sandage \& Tammann, 1985; Ferguson 1989).  
While populations of LMCGs seem to follow basic dwarf galaxy
photometric scaling laws,
such as surface brightness with magnitude, there is relatively 
little detailed information about their intrinsic properties.  To further
understand these objects, and their origin, we
have begun a series of papers to address these issues.

In Conselice, Gallagher and Wyse (2001a and 2002) (hereafter Papers I
and II), we outlined several of the basic problems, and possible solutions, for
understanding the origin of LMCGs.  In Paper I we argued that Virgo cluster
LMCGs as an aggregate population have dynamical properties and are spatially 
distributed in a way consistent
with their being accreted after the development of the cluster core, as 
defined by the giant elliptical galaxies.  
This idea is supported by the high  dispersion and non-Gaussian 
distributions of LMCG radial velocities and the large range in 
projected spatial positions of Virgo LMCGs.
We also found that the velocity dispersion of the LMCGs exceeds that of 
the giant ellipticals in the cluster core by a factor consistent with the 
LMCGs being a weakly gravitationally bound and accreted component of 
the cluster.  

In Paper I, we also argued that LMCGs cannot originate from galaxies 
born as field dEs or in groups because of the high LMCG to giant 
galaxy ratio in the Virgo
cluster. If Virgo LMCGs originated solely from accreted field dwarfs we
would expect the same
giant-to-dwarf ratio in galaxy clusters as in the field.  However, 
more LMCGs are found per giant cluster galaxy in nearby rich clusters (e.g.,
Thompson \& Greggory 1993; Secker \& Harris 1996).  
Paper I further argues that some Virgo LMCGs could originate from a more 
massive precursor population, such as disk galaxies, an idea also 
suggested in N-body models (e.g., Moore et al. 1998).

Paper II is the observational basis of the present paper, and in it we discuss 
methods for distinguishing different Perseus cluster galaxy populations.   
Paper II presented the observational results of a 173 arcmin$^{2}$ optical
survey of the central region of the Perseus cluster 
(v = 5366 \kms, D = 77 Mpc).   We argued, using the
data in Paper II, that different galaxy populations can be separated from
each other and background galaxies by their photometric and structural 
properties.  We also demonstrated that 
for faint M$_{\rm B} > -15$ Perseus cluster LMCGs, the universal elliptical 
galaxy color-magnitude relationship (CMR) breaks down.  
Previously (e.g., Bower, Lucey \& Ellis 1992) is has been thought
that the CMR was a linear relationship between magnitude and color.
However, this is not the case at faint, M$_{\rm B} > -15$, magnitudes, 
due to the large range in colors found for faint
cluster galaxies at a given magnitude (Paper II).  Here we study this 
deviation from the color-magnitude
relationship in detail to help determine the origin of LMCGs.

In this paper (Paper III of the series) we further
explore in detail the properties of LMCGs in Perseus and in other nearby
clusters, and compare their stellar-population characteristics to model 
predictions.  In particular, we examine whether or not the dispersion in the 
Perseus cluster
color-magnitude relationship for faint LMCGs with M$_{\rm B} > -15$ can be 
reproduced in various models such as: formation of LMCGs from stripped spirals 
born in the field, 
self-enrichment induced by the confinement of metals in LMCG halos
from the intracluster medium, and in situ formation out of intracluster
gas.  We also examine how these objects compare with predictions 
of various CDM models in which the mass and stellar populations of low-mass 
galaxies formed 
early, $> 10$ Gyrs ago (e.g., Dekel \& Silk 1986), and had little, to no, 
active mass evolution or star formation since.   Our general conclusion is
that a significant proportion of LMCGs are consistent with originating
from stripped higher mass galaxies accreted into clusters.

This paper is organized as follows:
in \S 2 we discuss the observational data and selection criteria for
LMCGs,  presented in
detail in Paper II, \S 3 gives an analysis of the early-type LMCG properties,
while \S 4 compares
these observations to various theories, and \S 5 gives a summary of our
findings.  We assume a Perseus distance of 77 Mpc throughout this paper,
giving a scale of $\sim$20~kpc per arcmin.

\section{Observations and Data}

\subsection{Observations}

All of the Perseus cluster imaging data used in this paper were taken with
the WIYN 3.5m f/6.2
telescope at the Kitt Peak National Observatory.  A detailed
description of the data and a basic analysis of the photometric
and structural properties of the cluster members, including the 
Perseus cluster luminosity function, was presented in Paper II.   Briefly,
the data come from two CCD cameras on WIYN, the S2kB and Mini-Mosaic.
The S2kB is a thinned
2048$^{2}$ pixel charged coupled device (CCD).  The scale of the
images is 0.2\arcsec\, pixel$^{-1}$, with a field of view 6.8\arcmin
$\times$ 6.8\arcmin.   The S2kB imaging used to obtain deep B and R photometry
took place on the nights on 1998 Nov 14 and 15 over an area of
 $\sim$173 arcmin$^{2}$, in photometric conditions.
The average seeing for all images over the two nights was 0.7\arcsec.
Mini-Mosaic U-band data taken in 1999 December were also photometric with
image fields 9.8\arcmin\, on a side.  Only a fraction of the galaxies
with BR photometry have reliable U-band data due to the small field covered
with the Mini-Mosaic camera and defective areas on the mosaic chips due
to imperfections and reflections.  Our observed area was chosen to cover the 
central part of Perseus near NGC~1275, and along the linear array of large
ellipticals to the west of NGC~1275 (see Paper II).

To calibrate the photometry, we fitted zero point offsets, airmass and
color terms using UBR Landolt standard star photometry, acquired
during each night (Paper II).  Photometry for each galaxy was done in
two ways.  Magnitudes representative of the entire galaxy were
measured within an aperture, R$_p$, equal to three times an inverted
Petrosian (1976) radius, or R$_{p}$ = 3 $\times$ r($\eta$ = 0.5).  We
also measured core magnitudes and colors within the central 2\arcsec\, of
each galaxy.  We also correct the photometry for Galactic extinction,
and for a slight k-correction. For more details on the sample
selection and the data themselves, see Paper II.  Monte Carlo
simulations, and comparisons between the S2kB and Mini-Mosaic
photometry, which is further discussed in Paper II, also argues that our
photometry is reliable down to B = 24 or M$_{\rm B} = -10.5$ at the adopted
distance of the Perseus cluster.

\subsection{Early-type LMCG Selection and Definition}

This section briefly discusses how we are able to confidently identify LMCGs
in the Perseus cluster from other cluster member and background systems
in a reliable manner.  See Paper II for a more detailed discussion
and description of these issues.  The summary of our selection criteria is 
the following:

\noindent (i). Total $(B-R)_{0}$ colors of  $<$ 2.  Galaxies redder than
this are almost always in the background. From stellar population
synthesis models (Worthey 1994), no normal passively evolving galaxy
with metallicity [Fe/H] = +0.5 is redder than $(B-R)_{0}$ $\sim$ 2 after an 18
Gyr initial burst of star formation. We further only investigate in detail
Perseus LMCGs with magnitudes M$_{\rm B} < -12.5$ to be more certain of 
cluster membership and to have very accurately measured 
photometric and structural parameters.

\noindent (ii). Symmetric, round or elliptical shapes, without
evidence for internal structures that might be due to star
formation, spiral arms, or other internal features.  Dwarf 
ellipticals/spheroidals appear this way, whereas 
background galaxies are often morphologically disturbed and
can be identified as such in high resolution images (e.g., Conselice 2001).

\noindent (iii). A central surface brightness in a 2\arcsec\, aperture fainter
than $\mu_{{\rm B}}$ = 24.0 mag arcsec$^{-2}$ and a non-centrally concentrated
light profile that is close to exponential (see Paper II and \S3.1).  
These are properties
of nearby dwarf ellipticals and can be used to distinguish
LMCGs from giant ellipticals and background systems.

\noindent (iv) As discussed in Paper II, we 
also limit our study to objects that are detected above a 5$\sigma$ 
confidence. 

Figure~1 shows Harris R-band images of the LMCGs which we study in this paper
and which constitute the galaxies that make up the low-luminosity scatter
in the Perseus color-magnitude relationship (Paper II, Figure~2 \& \S 3.4).    
We also some-what arbitrarily separate LMCGs into blue and red types for 
discussion purposes.  We define blue Perseus galaxies as 
those which are within $+2 \sigma$ of the color-magnitude
relationship scatter at M$_{\rm B} = -17$ on the red side and everything
bluer than this (see Conselice 2002).  LMCGs which are $>2\sigma$ redder than 
the CMR are denoted as red systems.  Figure~3 show the $(B-R)_{0}$ color 
histogram for Perseus LMCGs, down to M$_{\rm B} = -11$,
with the red and blue systems shaded differently (see Paper II). 

In addition to Figure~1, which displays the R-band images of
the 53 galaxies studied in this paper, Table~1 lists their positions,
photometric and structural properties. For reference,  
the absolute magnitude
M$_{{\rm B}}$, $(B-R)_{0}$ color, and identification number are also listed in 
Table 1 and printed on Figure~1 for each galaxy.
 
\section{Results}

\subsection{Structures and Surface Photometry of the LMCGs}

\subsubsection{Previous Work}

In Paper II we studied the basic photometric and structural properties of
the Perseus LMCGs, including simple global morphological
indexes such as asymmetry and concentration (Bershady, Jangren \&
Conselice 2000; Conselice, Bershady \& Jangren 2000).   We found that the 
concentration indexes and asymmetries of the LMCGs are consistent with
dwarf elliptical-like objects.  We also found a strong correlation
between the central surface brightness and magnitudes of these objects,
with the same relationship seen for nearby dwarfs (e.g., Binggeli \&
Cameron 1991).   This suggests that all Perseus LMCGs
are objects that would be classified, if nearby, as dwarf elliptical or 
spheroidals and are also thus likely cluster members.  We go beyond
this basic examination in this paper to determine the detailed structural 
properties of the Perseus LMCGs.

\subsubsection{Surface Photometry}

Previous studies have found that LMCGs in other clusters, as well as 
Local Group dSphs, have exponential profiles, while
large ellipticals have steeper r$^{1/4}$ de Vaucouleur profiles and
are found to be more rounder than LMCGs (Ryden \& Terndrup 1994). 
What is the case for the Perseus LMCGs? To answer this, we 
perform surface photometry on all 53 early-type LMCGs and all the elliptical 
galaxies used in this paper.  
We fit elliptical isophotes to each galaxy image using the Fourier 
prescription of Jedrzejewski (1987), implemented by
the ELLIPSE routine in STSDAS.  We allow the ellipticity, position
angle, and center to change while fitting.  The general form
of the Fourier decomposition of intensity distributions is

\begin{equation}
I(\phi, r) = I_{0} + \sum_{i=i}^{4} (a_{i} {\rm sin}(i \phi) + b_{i} {\rm cos}(i \phi)),
\end{equation}

\noindent with $a_{i}$ and $b_{i}$ the fitted Fourier components, and
$\phi$ the
position angle at each fitted semi-major
length $r$. For each surface brightness profile,
$I(r)$, a general S\'{e}rsic profile of the form:

\begin{equation}
I(r) = I_{0} \times {\rm exp} (-(r/{\rm r_{0}})^{(1/{\rm n})}),
\end{equation}

\noindent is fit, where $I(r)$ is the intensity of an isophote at $r$, 
$I_{0}$ is the central intensity, and r$_{0}$ is the scale length.
The value n = 1 is for pure exponential profiles, and n = 4 is the de
Vaucouleurs
profile that fits giant elliptical galaxies fairly well (e.g.,
Gavazzi et al. 2000).
Table 1 lists the fitted profile parameters, n and r$_{0}$, for each LMCG.

Figure~4 shows the detailed surface photometry and fitted parameters for the
brightest, M$_{{\rm B}}$ $< -13.5$, early-type LMCGs in the Perseus cluster.
The other LMCGs are generally
too faint to obtain reliable measurements over a considerable range in radius.
The first panel shows the surface brightness profiles for these galaxies
and their best S\'{e}rsic profiles.  The fits are generally good at the
cores, but are sometimes poorer at larger radii.

The S\'{e}rsic fits for the galaxies listed in Table 1 are probably minimally
affected by resolution. We determined this
by fitting isophotes and S\'{e}rsic profiles to stars in the various
Perseus fields. Depending on the seeing of the images, the S\'{e}rsic index n for
these stellar fits
varies from 0.5 - 0.6, which is among the lowest values found for the LMCGs.
The S\'{e}rsic radius r$_{0}$ is also very small $\sim$0.5\arcsec, lower
than all but a few of the LMCG values.

\subsubsection{Isophotal Fits}

The second panel in Figure~4 shows the fitted ellipticities, 
$\epsilon = (1 - a/b)$, for each isophote as a function of radius.  The 
average ellipticity within
5\arcsec\, for each object is also printed in the second panel.  In
general, the ellipticities are constant with radius and most early-type
LMCGs are
fairly round with $\epsilon$ $< 0.3$, with an average ellipticity within
5\arcsec\, for the objects in Figure~4 equal to 0.26$\pm0.16$ (Table 1 also 
lists these $<\epsilon>$ values).  Other studies demonstrate that Virgo dwarf
ellipticals are generally flatter than giant Es (Ryden \& Terndrup
1994; Ryden et al. 1999).  A few Perseus LMCGs have ellipticities that
systematically increase at larger radii, i.e. are
flatter in their outer parts.  This does not significantly
differ from the average ellipticity of giant ellipticals, $<\epsilon>$ =
0.24$\pm$0.11 (from data in Peletier et al. 1990 and Ryden et
al. 1999).

The position angles (P.A.) of the isophotal fits (column 3 on Figure~4)
are relatively constant with radius
for all galaxies, indicating that isophotal
twists are not common in Perseus LMCGs.  The fourth column shows
the a$_{4}$ component to the Fourier fits.  The value of a$_{4}$ 
indicates the degree
of deviation from purely elliptical isophotes, with a$_{4} > 0$ indicating
disky isophotes, and a$_{4} < 0$ for boxy isophotes. The
average a$_{4}$ for each galaxy is printed in the fourth panel of Figure~4,
with nine LMCGs having a$_{4} < 0$ and five with a$_{4} > 0$.   Most
of the LMCGs have very small deviations from a$_{4}$ = 0, and only
two LMCGs (31900.4+4129, 31941.7+4129) have large a$_{4}$ values.  These  
a$_{4}$ values are roughly
consistent with elliptical isophotes, with small variations, as is also seen
for Virgo cluster dEs (Ryden et al. 1999).

\subsubsection{Basic Interpretation}

Most Local Group dwarf-ellipticals have observed exponential profiles, or 
King profiles (Faber \& Lin 1983).  The
more general S\'{e}rsic profile is also found to provide a 
good fit to dE profiles, with most
indexes at n $<$ 2 (Davies et al. 1988; Binggeli \& Cameron 1991). 
We find that the brightest galaxies we classified as giant ellipticals
are well fit by de Vaucouleur profiles, although some of the fainter
ellipticals appear to be better fit by an exponential (Figure~5). 

Some previous studies found that more luminous Virgo dEs have surface 
brightness profiles with n$>1$ (e.g., Young \& Currie 1994; Durrell 1997).
Table 1 and Figures 4 and 5 shows
that most of the early-type LMCGs, chosen by morphology and color as
cluster members, have nearly exponential profiles similar to dwarf elliptical
galaxies. Only two LMCGs have higher n values near n $\sim 3$, although
classified dEs in other clusters can have S\'{e}rsic indexes this high 
(Binggeli \& Jerjen 1998).

We find no strong correlation between n and absolute magnitude for
galaxies at M$_{\rm B} > -17$ (Figure~5),
contrasting with earlier claims for dE galaxies in the Virgo cluster 
(e.g., Young \& Currie
1994; cf. Binggeli \& Jerjen 1998). In general it appears from Figure~5 that 
the brightest galaxies have the highest S\'{e}rsic indexes, n, although this 
correlation does not extend to fainter objects.    It is however important to 
note that if there are very compact LMCGs with very high S\'{e}rsic n values, 
then we are
liable to miss them in our morphological selection, as these objects will 
masquerade as stars or background ellipticals.

There is also no strong correlation between n and the $(B-R)_{0}$ color of
the LMCGs (Figure~5).  If these LMCGs were contaminated by smooth
symmetric background galaxies (the only ones that would pass our morphological
criteria), we would expect them to be red objects with steep, non-exponential,
surface brightness profiles.  No red object chosen as a LMCG was found to
have this characteristic.

\subsection{Internal Color Structures of LMCGs}

A first impression of the Perseus LMCG images (Figure~1) is that 
some are nucleated; based
solely on their structural appearance 17 (32\%) have evidence of an
excess central brightness. 
A critical observational test for determining the origin of
early-type LMCGs, and these nuclei, is whether or not any color
gradients exist within these galaxies.  Pronounced radial color gradients 
are infrequently seen in dE 
galaxies (e.g., Caldwell \& Bothun 1987; Vader et al. 1988; Cellone et al.
1994; Durrell 1997) and we find the same here. 
A lack of systematic color gradients in Perseus early-type LMCGs can be
demonstrated in several ways, consistent with our later use of single stellar
population synthesis models.
First, color maps of bright dEs (Figure~6) reveal that these 
early-type LMCGs have
no significant color gradients, or core colors that differ from those of
the underlying light. 

We further assessed the importance of radial 
color gradients by comparing the measured color of
the core (central 2$\arcsec$) for each dE to its total color, which
includes the core regions.
Figure~7 plots this relationship, where the solid squares are the color
differences in the
nucleated early-type LMCGs, and the open circles are the color differences
in the non-nucleated objects.
There is no significant color difference between the two different types, and 
there is no systematic trend between this color difference as a function of
color.  The color
differences $(B-R)_{0}$$^{{\rm tot}} -$ $(B-R)_{0}$$^{{\rm core}}$ and 
the 1$\sigma$
variations are essentially the same for the two
populations: -0.05$\pm$0.16 for the nucleated LMCGs and -0.09$\pm$0.17 for
non-nucleated ones.  This difference, 0.04 mag, is within our photometric
color error (Paper II).   The average colors ($\pm$1$\sigma$) of the various 
components of these systems are also essentially the same with $<(B-R)_{0}> =$
1.23$\pm$0.32 for integrated LMCGs colors and $<(B-R)_{0}> =$1.29$\pm$0.33 for
their cores.

The lack of color gradients can also be demonstrated 
by measuring the difference in colors between the 
outer regions of the LMCGs, which does not include their cores, and 
their core colors.  We apply this test by dividing galaxies into their
inner parts ($r=$ 2\arcsec) and the rest of the galaxies' light at 
r $>2$\arcsec.  The results of this comparison are 
shown by the solid triangles on Figure~7.  Again, we 
find few Perseus LMCGs with significant color gradients between the centers
and outer regions.
The few LMCGs that do show significant differences could be
galaxies with stellar populations at different ages and metallicities in their
cores and envelopes.  Some non-nucleated LMCGs also have red cores, possibly
the result of dust, very old stellar populations, or accreted
globular clusters (Lotz et al. 2001).  A few dEs in the Virgo cluster also
contain blue nuclei, but the majority of nucleated objects
have a central color similar
to their host galaxies (Durrell 1997). Further observations,
especially spectroscopy, will be necessary to determine if recent star
formation could have produced the bluer nuclei.

\subsection{Surface Densities and the Dwarf to Giant Ratio}

The early-type
dwarf to giant ratio (EDGR; Secker \& Harris 1996; Phillipps et al. 1998) 
provides a quantitative
measure of the form of the luminosity function, where galaxies are 
selected both by luminosity and morphology (Driver, Couch, \&
Phillipps 1998).
In our 173 arcmin$^{2}$ survey of the Perseus cluster, 
we find 160 candidate early-type dwarf
elliptical-like LMCGs with  
M$_{\rm B} < -11$ (Paper II), giving a
surface number density of $\sim$1 arcmin$^{-2}$ or $\sim$2000 early-type
LMCGs Mpc$^{-2}$.  The giant galaxy density is $\sim$260 giants 
Mpc$^{-2}$.
This gives an EDGR ratio of $\sim$8, which is lower than the Virgo or
Coma value at the brighter magnitude limit M$_{\rm B} = -12.5$ (e.g.,
Ferguson \& Sandage 1991; Secker \& Harris 1996).
For LMCGs with M$_{\rm B} < -12.5$, we find a surface
density of $\sim$0.5 arcmin$^{-2}$ or $\sim$1000 Mpc$^{-2}$.  This gives
an EDGR ratio of $\sim$4, compared with the values 9.4 and 9.3 computed
in the Coma and Virgo clusters (Secker \& Harris 1996).  Table~2 lists
the values of the EDGR computed for Perseus and values from the 
Virgo and Coma clusters. 

This result suggests that the EDGR ratio at the center of the 
Perseus cluster is lower than published values for the less dense 
Virgo cluster and integrated value for the Coma cluster 
(Ferguson \& Sandage 1991; Secker \& Harris 1996).
The EDGR ratio was found by Ferguson \& Sandage (1991) to correlate with
galaxy group density, with denser groups displaying higher
EDGRs.  Other studies suggest an opposite trend for rich clusters, such
that denser cluster regions contain lower EDGR ratios (Phillipps et al. 1998).
A reduced surface density of LMCGs near the center of a rich cluster
has been noted before (Thompson \& Gregory 1993; 
Driver et al. 1998; Adami et al. 2000),
and is possibly the result of galaxy destruction rather than initial
conditions (Conselice 2002; \S 4.3).  As pointed out by Thompson \& Gregory 
(1993) in the Coma cluster,
the processes creating LMCGs from stripping could become 
so efficient in 
the cores of rich clusters that many LMCGs are destroyed, or disrupted to 
the point where they are too faint to
be detected.

\subsection{The Color-Magnitude Relationship: Metallicity, Ages and Mass
to Light Ratios}

In Paper II we argued that for Perseus LMCGs fainter than M$_{\rm B} 
\sim -15$, 
there is an increase in the scatter about the universal linear CMR established
for giant ellipticals (Bower et al. 1992) (see Figure~2).   This increase
cannot be accounted for by photometric errors for systems more 
luminous than M$_{{\rm B}} \sim -12.5$ (Paper II).   Evidence that faint 
cluster dE galaxies scatter significantly from a linear CMR have
been available for some time (e.g., Caldwell \& Bothun 1987), but this has yet 
to be fully characterized, explained, or appreciated. 
To understand and characterize the galaxies that make up 
this scatter, we limit our detailed analysis to the sample of 
early-type Perseus LMCGs with M$_{\rm B} < -12.5$ 
listed in Table 1 and displayed in Figure~1.  

The origin of the relationship between the colors and magnitudes of
giant cluster elliptical galaxies has been debated since its initial
characterization by Sandage \& Visvanathan (1978), but it 
is now generally regarded as a relationship between
a galaxy's mass (traced by its magnitude) and metallicity (traced by
its color), with the stars in each galaxy at similarly old ages 
(Arimoto \& Yoshii 1987;
Bower et al. 1992).   This idea is supported through studies of
cluster galaxies at higher redshifts, where the scatter in the 
color-magnitude relationship remains low (Ellis et al. 1997; 
Stanford, Eisenhardt \& Dickinson 1997).   Metallicity as a driver of
cluster galaxy colors is also an inherent aspect of various formation 
models (Arimoto \&
Yoshii 1987; Kauffmann \& Charlot 1998).   Below we demonstrate
what our broadband UBR colors tell us about the nature of the stellar
populations in LMCGs; specifically we demonstrate that the red colors
of some LMCGs imply that they must be particularly metal enriched and that this
enhanced enrichment is at least partially the cause of the large scatter of 
these galaxies from the color-magnitude relationship.

\subsubsection{UBR Color-Color Plots: Constraining Ages and Metallicities}

As we argue in \S 3.2 the internal colors of Perseus LMCGs are nearly 
homogeneous, such that we can to a first approximation consider their stars 
as single stellar 
populations (SSPs), to place constraints on their ages, metallicities and 
stellar masses.  We investigate the properties of these stellar populations
by comparing spectral energy distributions of LMCGs to the stellar population 
synthesis models of Worthey (1994), including the Padova isochrone library 
(Bertelli et al. 1994).

Figure~8 shows model $(U-B)_{0} - (B-R)_{0}$ color-color tracks at three
different ages ($\tau = 18, 12, 5$ Gyrs), plotted as a function of 
metallicity, with each panel showing a different age, 
along with the Perseus cluster data.  All galaxies fall 
along, or close to these Worthey (1994) isochrones.
If Perseus cluster galaxies contain stellar populations formed at roughly
the same time, and are old, as clusters ellipticals are thought
to be (e.g., Kuntschner \& Davies
1998; Trager et al. 2000),  then Perseus giant ellipticals have near solar 
metallicity stellar 
populations, while the Perseus LMCGs consist of stars with a range of
lower metallicities.   

Figure~9 shows corresponding color-color diagrams of models at three constant 
metallicities at different ages, with each panel containing a different
constant metallicity model.  From these figures it is clear that the 
colors of Perseus galaxies can be accounted for by a mix of various ages as
and metallicities.  However, combining information from
Figures~8 and 9 allows us to place some constraints on the ages and 
metallicities
of the stellar populations that make up the Perseus LMCGs.  Figure~9
shows that the [Fe/H] = -0.5, and any lower metallicity models, are 
unable to account for the reddest
LMCGs, even if the ages of the stellar populations are $\sim 18$ Gyrs old, 
which is unlikely given the currently favored $\sim$~13 Gyr age of the 
universe (e.g., Ferreras, Melchiorri \& Silk 2001). What
this means is that the elliptical galaxies, and the reddest LMCGs with
$(B-R)_{0}$ $> 1.4$, must have metallicities with [Fe/H] $> -0.5$, unless
these systems are dusty.  If the stars
in the reddest LMCGs are younger than $\sim 12$ Gyrs, and dust free, 
their metallicities
must be even higher than [Fe/H] = $-0.5$.   Due to an
age-metallicity degeneracy in our broad-band colors this is the only
constraint we can place on the metallicities and ages of the red LMCGs.
We can place some constraints on the ages of the blue LMCGs by also using 
Figures~8 and 9.  If the metallicities
of the stellar populations in the bluest LMCGs, at $(B-R)_{0}$ $\sim 1$, are 
solar, [Fe/H]
= 0, then their ages cannot be less than $\sim 1$ Gyr old.  If these 
blue LMCGs have metallicities lower than solar, as seems likely given
spectroscopic and photometric results demonstrating that blue
LMCGs in other clusters are metal poor (e.g., Held \& Mould 1994; 
Rakos et al. 2001), then their stars must be older than 1 Gyr.

Based on these arguments we will often 
assume that the $(B-R)_{0}$ colors of the LMCGs
and giant ellipticals is representative of the metallicities of their
stellar populations.  This is not a new idea, as it is now generally 
thought that the colors of early-type galaxies in clusters are dominated by
metallicity as opposed to age (e.g., Kuntschner \& Davies 1998; 
Vazdekis et al. 2001).  Prior observations through spectroscopy and
narrow-band imaging also suggest that low-mass cluster galaxies
have spectral signatures which are dominated by metallicity  (e.g., Thuan 
1985; Bothun \& Mould 1988; Held
\& Mould 1994; Gorgas et al.  1997; Poggianti et al. 2001). 

\subsubsection{Mass-Metallicity Diagram and Mass to Light Ratios}

Colors and magnitudes are observational quantities that we can
convert into metallicity and mass given our assumptions in
\S 3.4.1.  By assuming that the $(B-R)_{0}$ colors of LMCGs and giant
ellipticals trace metallicity in an old stellar population, we can find
[Fe/H] using the relationship from
Harris (1996) for globular cluster systems,
dichotomy remnants telescope
\begin{equation}
{\rm [Fe/H]} = 3.44 \times (B-R)_{0} - 5.35.  
\end{equation}

\noindent Transforming the magnitudes of objects into stellar masses 
($M_{\rm stellar}$) requires knowledge of the stellar mass to light
ratio ($M_{\rm stellar}/L$), which is a function of
metallicity [Fe/H] and age ($\tau$),

\begin{equation}
M_{\rm stellar} = \frac{M_{\rm stellar}}{L_{B}}({\rm [Fe/H]}, \tau) \times L_{B} 
\end{equation}
$$= \frac{M_{\rm stellar}}{L_{B}}([{\rm Fe/H}], \tau) \times 10^{0.4(5.45 - {\rm M_{\rm B}}).}$$

\noindent By assuming that metallicity correlates with $(B-R)_{0}$ as given in 
equation (3), and by adopting an age ($\tau$) for a stellar 
population, we can
determine the relationship between $(B-R)_{0}$ and (M$_{\rm stellar}/L$).  
We take the value $\tau = 12$ Gyr as a representative age, although
other ages do not give significantly different results in what follows.
In this situation we can then write (M$_{\rm stellar}/L$) as a power law of 
(B$-$R) such that, 

\begin{equation}
\frac{M_{\rm stellar}}{L}~\alpha~{\rm (B-R)_{0}}^\kappa.  
\end{equation}

\noindent The relationship
between (M$_{\rm stellar}/L$) and $(B-R)_{0}$ is plotted in Figure 10 for $\tau 
=$ 8, 12 and 14 Gyrs.  At $\tau = 12$ Gyrs, $\kappa = 2.4$, which 
we use to calculate the stellar mass of each Perseus galaxy from its 
luminosity.

Using equations (3) and (4) we can create a metallicity-stellar mass diagram 
(Figure~11).    Figure~11 shows that there are two stellar mass-metallicity
sequences in the Perseus cluster.  The blue (and now low-metallicity)
LMCGs appear to blend naturally into the low-mass sequence of
the higher mass elliptical galaxies.  However, part of the 
LMCGs form a separate sequence that includes
all the red LMCGs.    Based solely on this diagram it is unclear what
these two sequences imply for the formation and evolution of LMCGs, but it
is a curious feature that any proposed LMCG origin scenario must explain.

\section{The Origin of LMCGs}

\subsection{Primordial Galaxies}
 
A common theoretical presumption about low-mass 
galaxies, that we explore here, is that they
are very old objects that originated in their present forms soon after
the universe began, and possibly formed many of their stars
before reionization at $z >
6.5$.  Due to suppression of star formation caused by
photoionization, star-formation in low mass galaxies with old stars
should occur before reionization (Efstathiou 1992).
The low luminosities, and hence likely low masses, of the early-type 
LMCGs suggests they fit the basic criteria for old galaxies, as
they are found in dense regions, where hierarchical clustering
models predict the oldest galaxies should be most abundant (e.g., White \& 
Springel 2000). 

The most generally accepted idea is that dwarf galaxies form
from the gravitational collapse of baryonic material, possibly within 
dark matter halos, to produce the first
generation of stars, with the remaining gas removed through
supernova driven winds (e.g., Larson 1974; 
Dekel \& Silk 1986; Silk, Wyse \& Shields 1987;
Ferrara \& Tolstoy 2000).   Predictions of
detailed models show that dwarfs should be low metallicity systems with
minor color gradients and less clustered than giant galaxies.  
Low metallicity is a general feature for the formation of low-mass
galaxies in the early universe and we investigate this scenario here using
the data described in \S 3.4 as well as other photometric results 
(Rakos et al. 2001) supplemented by dynamical data (Pedraz et al. 2002; 
Geha, Guhathakurta \& van der Marel 2002) from other studies.

\subsubsection{Internal Dynamics}

In addition to predicting low metallicity systems, Dekel \& Silk (1986) 
show, based on their model assumptions, how the internal velocities, 
$\sigma$, of low-mass galaxies
should correlate with $M_{\rm total}/L$ ratios and luminosities ($L$), 
such that $M_{\rm total}/L$
$\sim L^{-0.37}$ and $\sigma \sim L^{0.19}$.  We test these predictions
with Virgo cluster LMCG internal velocity measurements made by 
Pedraz et al. (2002)
and Geha et al. (2002).    The relationship between M$_{V}$ and
internal velocity $\sigma$ for these galaxies is shown in Figure~12.  When we 
fit
a power law to the relationship between $\sigma$ and L, we find
$\sigma \sim$ $L^{0.31\pm0.05}$.  The fitted exponent on $L$ is 
$\sim 2\sigma$ away from the
relationship predicted by Dekel \& Silk.  If we limit this fit to
the fainter LMCGs at M$_{V} > -17$  we find 
$\sigma \sim L^{0.18\pm0.12}$, in good agreement with the models
of Dekel \& Silk.  

After converting the sizes and velocity dispersions listed in Geha et al.
(2002) into a pseudo-total mass using the relationship $M_{\rm total} \approx
\epsilon {\rm r_{e}} \times \sigma^{2}$  we
find that the $M_{\rm total}/L_{V}$ ratios for these Virgo LMCGs increases at
lower luminosities.   Fitting these two quantities together we find
$M_{\rm total}/L_{V}$ $\sim L^{-0.25\pm0.24}$, broadly consistent with 
the Dekel \& Silk slope predictions.  
In general, the above arguments suggest that the limited number of LMCGs 
with measured internal
velocities are consistent with having evolved through a self-enrichment
supernova wind outflow model, such as that proposed by Dekel \& Silk, although
these internal velocity samples are dominated only by a few bright LMCGs that 
tend to follow giant galaxy scaling relationships.

\subsubsection{Stellar Populations}

The most complete previous analysis of the stellar populations for a sizable
population of LMCGs is that of Rakos et al. (2001), who estimated the
ages and metallicities of 27 Fornax cluster dEs.  Rakos et
al. (2001) find that there is a broad age and metallicity distribution for
this sample of Fornax dEs, with [Fe/H] values ranging from -1.6 to -0.4 and 
ages from 9 to 12 Gyrs.    From this, and the analysis in \S 3.4 where we 
conclude
that red Perseus LMCGs can not be metal poor objects, it is difficult for all 
LMCGs to be primordial objects formed
through the e.g., Dekel \& Silk paradigm as
we would expect to have a population of objects with homogeneously
old and metal-poor stellar populations.

We also argue that not all LMCGs can form in a simple early collapse
by directly comparing our data to CDM simulations
that predict the properties of low-mass galaxies that all formed early
in the universe.  In Figure~13 we plot the
color-magnitude diagram for the Perseus galaxies together with the
luminosities and colors derived from predictions of low-mass 
galaxies from the
$\Lambda$CDM hydrodynamic models of Nagamine et al. (2001) which are 
based on simple
feedback mechanisms that do not take into account environmental effects.  The 
high stellar mass objects in these simulations, with M$_{\rm stellar}$ 
$>2\times10^{9}$ h$^{-1}$ \solm, have
a wide range in stellar ages and metals due to the hierarchical nature of
their formation over time.  The low-mass galaxies with
M$_{\rm stellar} < 2 \times 10^{8}$ h$^{-1}$ \solm, on the other
hand, experience only a brief epoch of star formation, with
few new stars produced after the universe is $\sim 2$ Gyrs old.  The
conversion between metallicity and color for the Nagamine et al. (2001)
simulated galaxies was achieved by adopting the
Milky Way globular cluster-based relation (eq. 3).

There is a general agreement between the simulated Nagamine et al.
(2001) and Perseus cluster 
galaxies properties,
except at the faint and bright ends of the luminosity function.  Bright
Perseus galaxies are redder than the simulated points, and these
$\Lambda$CDM models do not predict the observed population of red low-mass
galaxies.  It does, however, 
appear that the $\Lambda$CDM simulated objects, based on simple
feedback mechanisms, can reproduce the blue LMCGs as 
metal-poor and old galaxies.

\subsubsection{Extreme Dark Matter Halos}

If the correlation of color and magnitude for bright galaxies represents a 
total
mass-metallicity relationship and LMCGs are primordial and evolve through 
self-enrichment, then the fact that the color-magnitude
relation breaks down could be the result of varying total mass to light
($M_{\rm total}/L$) ratios.  That is, LMCGs could contain large amounts of
dark matter.    A very high total mass to light ratio could account
for objects that have high metallicities, but appear faint, such
as the red LMCGs in the supernova - self-enrichment scenario.   That
is, when metals are produced during star formation
in these dark-matter-dominated galaxies they are not ejected into the 
intracluster medium due to the large gravitational potential of the galaxy, 
which essentially traps the metals long enough for subsequent generations
of more metal rich stars to form. 

The likelihood of this scenario can be examined using the color-magnitude
and mass-metallicity relationships and
properties of the red LMCGs.   For LMCGs at $(B-R)_{0}$$\sim$ 1.4 to be at a 
similar location of the mass-metallicity/color-magnitude relationship as a 
brighter elliptical galaxy at the same color would require a total mass
$\sim 3$ magnitudes higher, and thus a total mass to light ratio of
$M_{\rm total}/L$ $\sim$ (2.5)$^{3}$ $\times (M_{\rm total}/L)_{\rm
ellipticals}$ $\sim$ 15 $\times (M_{\rm total}/L)_{\rm
ellipticals}$.   If we take the $M_{\rm total}/L$ values for
ellipticals computed using galaxy-galaxy lensing by McKay et al. (2002),
which are 425$\pm$51 and 221$\pm$26 in the Sloan Digitized Survey g' and r' 
bands, then the inferred LMCG $M_{\rm total}/L \sim 
6\times 10^{3}$.  This would require an enormous dark matter content and
is about 10 times larger than any single galaxy $M_{\rm total}/L$ 
yet derived (e.g., Mateo 1998; Kleyna et al. 2002).  However the 
intracluster medium in clusters could have stripped gas from low-mass
halos early through ram-pressure (Gunn \& Gott 1972) before significant
star formation and without removing any dark
matter, thereby producing a galaxy with a high $M_{\rm total}/L$ ratio.

The best way to determine if the LMCGs have these high mass to light
ratios would be to measure their internal velocity dispersions out to a
large radius (e.g., Kleyna et al. 2002).  This type of information does not 
yet exist for any
LMCGs.  In fact, the only velocity dispersion measurements are those
in the central parts of LMCGs (Pedraz et al. 2002; Geha et al. 2002) where
dark matter is not found to be present in large amounts.  Future observations 
through
radial velocities of dE globular clusters or velocity dispersion profiles
out to considerable distances are needed to determine total LMCG masses within
a relatively large radius. 

\subsection{Non-Self Enrichment Stochastic Formation or Intracluster
Medium Containment}

The wide range in inferred metallicities for LMCGs suggests that a scenario
different from the standard dark halo plus supernova wind blow out (Dekel
\& Silk 1986) is occurring during their formation.  There are two alternative
scenarios we discuss here which can potentially reproduce some observational
properties of LMCGs.  One is that some LMCGs have delayed formation 
times, possibly due to the ultraviolet background intensity at high redshifts 
that keeps the gas in LMCGs photoionized until $z \sim 1$ (Babul \& Rees 1992).
LMCGs would then form later out of enriched intracluster gas.
This would be a formation scenario similar to, but not identical with,
how globular clusters are formed (Fall \& Rees 1986) in galaxies
by the cooling of gas.
The other possibility is that the intracluster medium is able to contain the 
metal rich ejecta of supernova, and subsequent star formation inside the halos
of LMCGs uses this material to produce subsequent generations of
metal-rich stellar populations.  

The first scenario is very attractive as it mimics the same properties,
and the lack of correlations between mass or luminosity with metallicity,
as found for globular clusters (Djorgovski \& Meylan 1994), and could explain
the existence of the faint blue galaxies (Babul \& Ferguson 1996).  
This scenario
would require that gas in clusters be able to cool to form stars
at different times.  Alternatively, a similar
method of LMCG formation could occur due to the outflowing material from the 
active nucleus at the center of NGC~1275 (Conselice et al. 2001b) 
interacting with the intracluster medium (e.g., Natarajan, Sigurdsson \& 
Silk 1998).

The present day iron abundance in the Perseus cluster center is slightly
lower than solar using the most optimistic estimates (e.g., Ulmer
et al. 1987; Ponman et al. 1990; Molendi et al. 1998; Ezawa et al. 2001).
If red LMCGs formed from gas in the intracluster medium with these abundances, 
this would explain their relatively high metallicities, assuming that 
the Fe abundance was put into place early from supernovae.  This
scenario would also naturally explain why, on average,
LMCGs with younger stellar populations in Fornax have higher metallicities 
(Rakos et al. 2001).

A problem with this idea is that the material in the intracluster medium
of the Perseus cluster must be able to cool and form stars without the
presence of a dark matter halo and at different times.   For this to occur 
dissipatively the
cooling time must be shorter than the dynamical collapse time.  The
10$^{7}$ K gas at the center of the Perseus cluster cools in a Hubble
time, making it possible that some of this material goes into creating
LMCGs at the cluster center.  This leads to another problem  
in that LMCGs produced in this method would likely have radial 
velocities close to that
of the intracluster gas and would be located towards the center of the 
cluster, where
the gas cooling time is the lowest.  Dynamical processes such as 
viralization within the cluster occur a Hubble time or longer, and thus
these LMCGs would retain their birth velocities (see Paper I).  
The radial velocity and spatial distributions of LMCGs in clusters have
larger dispersions that the centrally located giants (Paper I), making 
formation from the intracluster medium unlikely. 

The possibility that the LMCGs do not lose their supernova 
products due to confinement by
the intracluster medium would occur if the kinetic energy of
supernova ejecta was not large enough to displace the intracluster gas
surrounding the halo of a LMCG, and as such this material fell back into
the LMCG where it was used in further generations of star formation (Babul
\& Rees 1992).
A simple interpretation of this scenario would however suggest that the
metallicities of the stellar populations in LMCGs should all be similarly
metal-rich, which is contrary to the wide range of computed metallicities
in Perseus (\S 3.4) and Fornax LMCGs (Rakos et al. 2001).   There are
also similar problems with this model reproducing the velocity structure
of LMCGs (Paper I).

\subsection{Mass Removal Processes}

Another method of forming LMCGs is through the tidal stripping of
higher-mass galaxies to form lower mass systems.  The occurrence, 
strength, and importance of this process in clusters has been debated for 
many years (e.g., Spitzer 1958; Gallagher \& Ostriker 1972; Richstone 1976; 
Strom \& Strom 1978; Merritt 1984; Aguilar \& White 1986; Thompson
\& Gregory 1993; Lopez-Cruz 1997; Moore et al. 1998; Gnedin 1999; 
Conselice 2002) but is likely occurring (e.g., Conselice \& Gallagher 1998,
1999).

As interactions should occur in clusters, it is possible that some
galaxies in clusters,
ones either present at its formation or ones that are later accreted, collide,
or become stripped of enough mass to mimic the morphological and structural
properties of LMCGs and dwarf ellipticals.  Stripped gaseous
material from disk galaxy mergers within the cluster could also develop and 
condense into low-mass galaxies as tidal dwarfs
(e.g., Duc \& Mirabel 1998).   Candidate tidal dwarf galaxies in nearby
clusters have observed high metallicities (Duc et al. 2001) that could 
evolve into red LMCGs.  As tidal dwarfs are produced through low speed
galaxy interactions, it is not clear if enough of these interactions would 
have occured
in the past to account for the red LMCG population seen in nearby clusters.  
Further observations and modeling of the
production and survival of tidal dwarfs in a dense cluster-like
environment are necessary to demonstrate the likelihood of this process 
creating substantial dwarfs, but it is one possible out come of the
interactions discussed here.

The strength of these interactions,
and whether they arise from the overall mass distribution 
of the cluster (including dark matter) or from individual galaxies through
high speed encounters depends on the fraction
of cluster mass bound to galaxies, an uncertain and hard to measure
quantity (Natarajan et al. 1998b).  As such, we examine both of these 
dynamical mass stripping scenarios below.

\subsubsection{Tidal Limitation}

When a galaxy enters the core of a cluster it encounters intense gravitational
tides from the cluster potential that can change the galaxy's size and mass.   
For galaxies whose orbits take them to within twice the core radius ($R_{c}$) 
there is a limiting
relationship between the truncated tidal radius ($r_{T}$) and the internal 
velocity dispersion $\sigma_{g}$ of the galaxy (Merritt 1984).   
The Perseus cluster has a core radius of 
11\arcmin\, (Kent \& Sargent 1983) and thus nearly all galaxies observed in 
our survey (Paper II)
are within 2$\times$$R_{c}$, minus any projection effects.   The ratio of
the tidal radius of a galaxy, $r_{T}$, to the core radius of a cluster, 
$R_{c}$, is given by (Merritt 1984),

\begin{equation}
\frac{r_{T}}{R_{c}} = \frac{1}{2}\times \frac{\sigma_{g}}{\sigma_{cl}},
\end{equation}

\noindent where $\sigma_{\rm cl}$ is the observed velocity dispersion of
the cluster, and $\sigma_{\rm cl}$ = 1260 \kms\, for Perseus (Kent \&
Sargent 1983).  Using these values, we find that a tidally limited
system has a ratio $r_{T}/\sigma_{g}$ = 0.13 kpc (km/s)$^{-1}$.  While
we do not have internal velocities for any of our Perseus LMCGs, we do
have measured sizes which we can use to determine internal velocity 
dispersions by assuming that these objects are at the tidal limit.  If we
use $n \times r_{e}$ as the tidal radius ($r_{T}$) then we find that
the internal velocity dispersion for Perseus LMCGs would vary from
$\sim n$ \kms\, to $8n$ \kms, and we would except $n \sim 2 - 3$.   
The derived $\sigma$ values are similar to, or just lower than,  the
measured internal
velocity dispersions of nearby dwarf ellipticals (Mateo 1998) and for
LMCGs in nearby clusters (Pedraz et al. 2002; Geha et al. 2002).  In the
Virgo cluster, the computed tidal radii for the six galaxies with internal
velocities studied by Geha et al. (2002) 
are larger by a factor of 5 - 7 than their effective radii, using the Virgo 
radial velocity given in Paper I and
the core radii computed by Binggeli, Tammann \& Sandage (1987). This
implies that LMCGs in Virgo and Perseus could be tidally limited systems.

\subsubsection{Collisional Stripping}

If a large fraction of the mass inside a cluster is attached to galaxies 
then collisional interactions between galaxies will be the dominant
method for removing mass.   Most encounters between 
galaxies in clusters occur at high speeds,
with $\delta$v $\sim$ $\sigma_{\rm Cluster}$  
$>>$ $\sigma_{g}$, which increases the internal energy in each 
system with a resulting gradual mass loss.  In this situation the
impulse approximation for the increase of internal energy of spherically
symmetric systems
can be used.  This approximation is found to be reliable in many
circumstances (e.g., Richstone 1976; Dekel et al. 1980; Aguilar \& White 1985; 
Moore et al. 1998) and it is possible to 
parameterize analytically the mass loss rate in terms of the impact parameter,
p, and the perturbed galaxy's half-light radius, r$_{\rm e}$ (e.g.,
Aguilar \& White 1985)

The Aguilar \& White (1985) formalism allows us to investigate the effects of 
mass-loss from high-speed galaxy encounters in the Perseus cluster core.
A typical galaxy
orbiting through Perseus will undergo impulsive interactions from all
galaxies, but close encounters are more efficient at producing mass loss
than distant ones.  Without knowing the detailed orbital structure of every 
galaxy it is impossible to determine exactly the extent of individual
interactions.  Some estimates can be made by using averages, and
understanding the limitations where impulse encounters are ineffective
at removing mass.   For a galaxy with an initial mass of M = 10$^{10}$~\solm 
and size r$_{e}$ = 3 kpc interacting
with a perturber of mass M$_{p}$ = 10$^{11}$~\solm, at a velocity of
interaction V = 1200 \kms\, Aguilar \& White (1985) gives a mass loss rate 
$\eta = \delta$M/M = 0.1 per interaction.  After n interactions
for every m cluster crossings, there are q=n$\times$m total interactions, and
the total mass of the galaxy after these interactions is, assuming that the 
change in mass produces no change in the interaction efficiency,

\begin{equation}
M_{f} = M_{0}(1-\eta)^{q}.
\end{equation}

\noindent Aguilar \& White show (e.g., their Fig. 8) that the parameters we 
use in our calculations
are in regime where galaxies should expand due to collisions. This implies 
that after one collision, the next one is likely to be at a somewhat smaller 
${\rm p/r_e}$ just because ${\rm r_e}$ has increased. On the other hand, 
$\beta$ could decrease   
because $(M \times {\rm r_e})^{1/2}$ increases in the regime that we are 
considering. 
The important factor is whether the post-collisional change in $\beta$ or 
${\rm p}$ is more important, which is not easy to calculate.

However, we can determine from Figure~5 in Aguilar \& White (1985), that the $\beta$ 
term is likely to dominate and the efficiency of mass-loss should 
decrease with each collision.   We estimate a limit by assuming there 
is no change in the mean impact parameter, ${\rm p}$. Thus 
we can obtain an upper bound to the mass loss after q collisions, 
but in reality the process would become self-limiting.  The modeled 
collisions in Aguilar \& White (1985) also take place in an 
otherwise empty universe. Additional factors in real clusters, 
including perturbations from distant galaxies and the tidal field 
of the overall cluster (e.g., Merritt 1984) and the clumpiness of the 
cluster mass distribution (Gnedin 1999), may act to enhance 
these mass loss rates. 
    
What is the expected mass loss evolution for a galaxy of initial mass
10$^{10}$ \solm introduced
in a Perseus like cluster as a function of time starting at $z = 1$, assuming 
the mass loss rate $\eta$ remains constant?  As a lower limit, we assume
one p=10 interaction per core crossing with a crossing time
of $\sim$0.5 Mpc / 1200 pc/Myr $\sim$ 0.42 Gyrs. In this scenario q
= $2.4~\times~t_{clu}$, where $t_{clu}$ is the total time in Gyrs the galaxy 
has been in the cluster, or the time the cluster has had a fixed potential.  
Using the above description this galaxy will have a mass of $\sim 10^{8}$
\solm at $z \sim 0$ (see also Conselice 2002).  Thus, it is possible to 
produce LMCGs from high mass galaxies through collisional stripping (see
also Moore et al. 1998).

\subsubsection{Stellar Populations and Properties}

Given the above discussion we investigate, in this section, the idea that 
LMCGs are remnants of accreted disk galaxies.  In Paper I we argued that LMCGs
in the Virgo cluster have kinematic and spatial signatures suggesting these
galaxies, or their progenitors, were accreted in the last few Gyrs. 
Depending on the details of the interactions, accreted galaxies 
could lose $ > $90\%, to all, of their initial mass after undergoing
these tidal processes.   The effectiveness of mass loss could also be increased
by gas removal processes in unevolved disk galaxies.
If mass loss occured in the past, it is a natural
explanation for the production of some LMCGs from more massive, and hence
redder progenitors. 

Since many bulges, and the centers of other high mass galaxies, 
have dominant metal-rich populations
(e.g., Wyse, Gilmore \& Franx 1997) and observations of LMCGs in various
nearby clusters
also reveal metal-rich signatures and similarly red colors (\S 3.4 and e.g., 
Bothun \& Mould 1988; Held \& Mould 1994; Rakos et al. 2001), it is
possible that Perseus LMCGs are consistent with
being the remnants of more massive
progenitors whose outer parts have been stripped away. 

We can further examine the likelihood of collisional and tidal limitation 
scenarios
by examining the S\'{e}rsic scale length (r$_{0}$) of the LMCGs as a function of 
color (Figure~14).  The
fact that the red LMCGs (triangles) all have small effective radii is 
another indication that at least the red LMCGs have undergone some type
of tidal limitation process, or collisional removal of mass.  These
red LMCGs also have a higher surface brightness than the bluer
LMCGs at a given luminosity (Paper II), further suggesting that the red LMCGs
are stripped down components of larger galaxies.

By comparing
the colors of bulges of 257 Sbc galaxies from Gadotti \& dos Anjos (2001)
and Peletier \& Balcells (1997) 
to the colors of LMCGs we can test if there is a relationship between 
bulges and LMCGs by comparing the distribution of their colors.
Figure 15 is the histogram of $(B-R)_{0}$ colors from these two studies.  
Gadotti \& dos Anjos give $(B-V)_{0}$ and $(U-B)_{0}$ bulge colors which we 
convert into $(B-R)_{0}$ colors by determining the
relationship between $(B-V)_{0}$ and $(B-R)_{0}$ based on Worthey 
(1994) population synthesis models.  The conversion we 
use assumes that $(B-V)_{0}$ color is
the result of different ages at a constant metallicity of [Fe/H] = -0.2, but 
the results do not change significantly if we
assume colors are the result of different metallicities.
The distribution of bulge colors from
Gadotti \& dos Anjos (2001) is shown in Figure 15 as an open histogram
while the bulge colors from early type galaxies from
Peletier \& Balcells (1997) are shown as the shaded histogram.  The 
distribution of bulge colors is similar to the color distribution of the
LMCGs (Figure~3).   

A prediction from the tidal stripping
scenario is that a substantial number of 
intracluster stars should exist. These would be the bulk of the 
stars, $>99$\%, that were lost to create each stripped LMCGs.   
The existence of these intracluster stars is suggested through several
pieces of observational evidence.  Between 25\% - 50\% of the light from 
galaxy clusters originates in intracluster regions 
(e.g., Melnick, White \& Hoessel 1977; Bernstein et al. 
1995; Feldmeier et al. 2002).  
Evolved stars are also found in the intracluster medium (e.g., Ferguson, 
Tanvir \& von Hippel 1998;  Feldmeier, Ciardullo \& Jacoby 1998;
Arnaboldi et al. 20001; Durrell et al. 2002) that could be remnants of
stripped material from LMCG precursors.  There are also several 
examples of debris  arcs of diffuse light (Trentham \& Mobasher 1998; Gregg 
\& West 1998; Calcaneo-Roldan et al. 2000) and red distorted galaxies in 
clusters (e.g., Conselice \& Gallagher 1999) 
that could be contemporary examples of this tidal stripping process 
at work.

We can place further limits on tidal stripping occurring in
clusters by using the results of intracluster light and planetary
nebula surveys (Arnaboldi et al. 2002; Durrell et al. 2002).  
The metallicities of red giant branch and asymptotic giant branch
stars found in the Virgo cluster intracluster medium are found to be 
-0.8$<$[Fe/H]$<$-0.2, consistent with having
originated in moderate luminosity galaxies (Durrell et al. 2002).   The
total luminosity of the intracluster light within 2\deg of M87 is 
$\sim 2 \times 10^{11}$ \soll (Durrell et al. 2002), and if this material
originates from tidally striped 
objects whose remnants we now tentatively identify as LMCGs, we can compute 
how much 
material, on average, each LMCG must have lost.    Using the Virgo Cluster 
Catalog from Binggeli, Sandage \& Tammann (1985) we
can calculate the number of classified dwarf ellipticals, nucleated
dwarf ellipticals and S0s within 2\deg of M87.  We only consider these three
populations as they are the only types in Virgo that have kinematic 
evidence of past accretion (Paper I).

There are 170 dEs, 148 dE,Ns and 14 S0s within this radius.  On average,
if all the dEs and dE,Ns are remnants of stripped galaxies then $\sim
0.6 \times 10^{9}$ \soll was lost by each object.  This number remains
unchanged if we consider S0s to have lost a similar amount of mass.
If we consider only the dEs or dE,Ns as
remnants of this process then 1.1 $\times 10^{9}$ \soll and 1.3  
$\times 10^{9}$ \soll was lost by each dE and dE,N, respectively.  This is
enough material to suggest that each LMCGs could have been a moderate sized 
disk galaxies in the past. 

\section{Summary and Conclusions}

In this paper, we study the photometric properties of a complete sample of 
candidate early-type
low-mass cluster galaxies (LMCGs) in the Perseus cluster to place
constraints on their origin.   Our major findings can be summarized as:

\noindent I. Dwarf elliptical-like LMCGs at $-12.5 >$ M$_{\rm B}$ $>
-15$ scatter from the color-magnitude relationship extrapolated
from giant elliptical galaxies, although the `mean' LMCG still follows
the color-magnitude relationship.  There are individual LMCGs that are both 
too blue
and too red for their magnitudes.   If we interpret these colors as
indications of metallicity, then there exists a range of stellar population
metallicities in early-type LMCGs in Perseus, similar to the 
broad metallicity
distributions inferred for dwarfs in other galaxy clusters (Rakos et al. 2001; 
Poggianti 2001).  A range of
ages could also be present; this is not ruled out by our data, and
has in fact been concluded by others (e.g., Held \& Mould 1994; Rakos et al. 
2001). 

\noindent II. Using photometric and kinematic results from this, and other
studies, we show that some Perseus LMCGs have colors and population color
dispersions that are generally consistent with models that include 
early star formation plus feedback (Larson 1974), including those based on 
CDM cosmological scenarios (Dekel \& Silk 1986; Nagamine et al. 2001).   
Trends between velocity dispersion, luminosity
and total mass to light ratios for LMCGs at M$_{\rm B} > -17$ are consistent 
with the self-enrichment supernova model for low-mass galaxies formation 
(Dekel \& Silk 1986), although dynamical measurements have only been
completed for a few galaxies in Virgo. Because of their possible low 
metallicities and old stellar population ages, some LMCGs are 
potentially among the first galaxies to form stars in the universe.

\noindent III. The large number of LMCGs that are redder than the mean
color-magnitude relationship cannot originate from a simple self-enrichment
evolutionary path of initial formation from collapse plus later feedback.  
Our analysis and others (e.g., Rakos et al. 2001) reveal that these 
redder LMCGs probably contain stellar 
populations with high metallicities, up to solar.  These objects
therefore must have formed in a manner fundamentally different than the 
majority of Local Group low mass galaxies.  We identify four possible 
scenarios for explaining
these systems: containment of metals in a halo by the intracluster medium,
delayed formation out of intracluster gas with non-self enrichment 
(e.g., like globular clusters), very massive dark halos with 
$M_{\rm total}/L_{B} \sim 6 \times 10^{3}$, and as the remnants of stripped 
spiral galaxies.  Various
observational evidence concerning LMCGs themselves, and intracluster light,
suggests that the most likely of these scenarios is that 
the redder LMCGs are produced by tidal stripping.

Another more unlikely formation scenario, we have not discussed, is that
initial mass functions for stars in LMCGs have extremely low 
upper mass limits or very steep faint end slopes.  This would allow these 
galaxies to contain a excess amount of low-mass stars which would result in 
their observed red colors.  However, there is no obvious physical 
mechanisms that would result in this situation and we do not consider 
this scenario likely.  To further suggest which of the above scenarios is 
occurring
to produce LMCGs will require further observations.  We argue however
that no single explanation can likely account for all LMCGs, and
that the formation mechanisms for low-mass galaxies in clusters is
fundamentally different than that for low-mass galaxies in groups.

We thank the staff of WIYN and NOAO for their support in obtaining the
observations presented here. We also thank Kentaro Nagamine, Renyue Cen
and Jeremiah Ostriker for their $\Lambda$CDM model results.  We especially
thank Linda Sparke whose carefully reading and criticisms of a previous
version of this paper resulted in a nearly complete rewrite. This research was 
supported
in part by the National Science Foundation (NSF) through grants AST-9803018
to the University of Wisconsin-Madison and AST-9804706 to Johns Hopkins
University. CJC acknowledges support from a Grant-In-Aid of Research
from Sigma Xi and the National Academy of Sciences (NAS) as well as a
Graduate Student Researchers Program (GSRP) Fellowship from NASA and the
Graduate Student Program at the Space Telescope Science Institute
(STScI).

\newpage

\begin{figure}
\plotfiddle{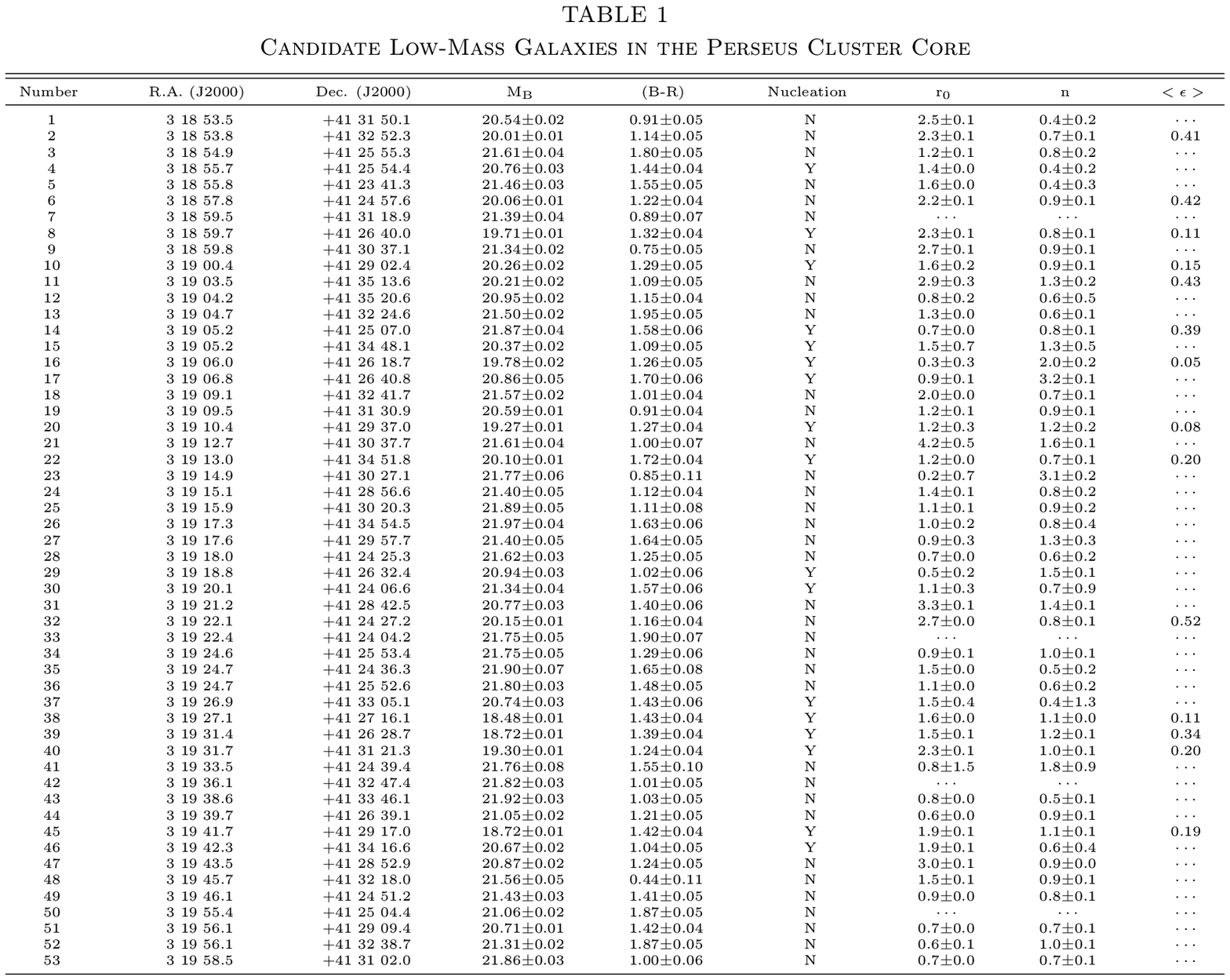}{6.0in}{0}{80}{100}{-265}{-175}
\end{figure}

\begin{figure}
\plotfiddle{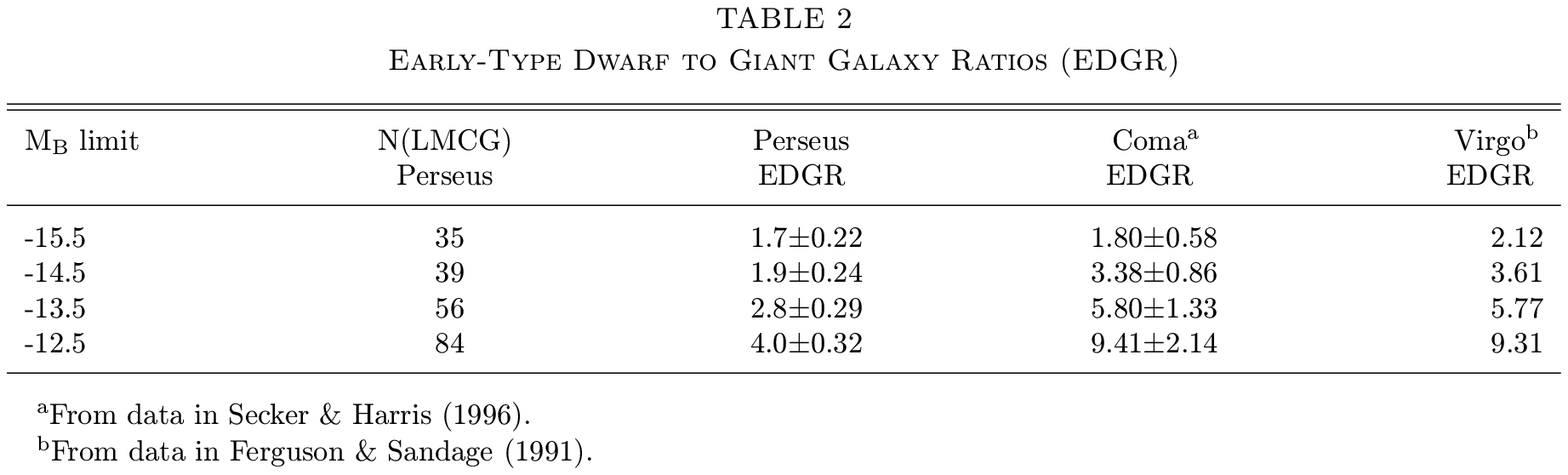}{6.0in}{0}{80}{100}{-265}{-175}
\end{figure}

\newpage

\setcounter{figure}{0}

\clearpage
\begin{figure} 
\plotfiddle{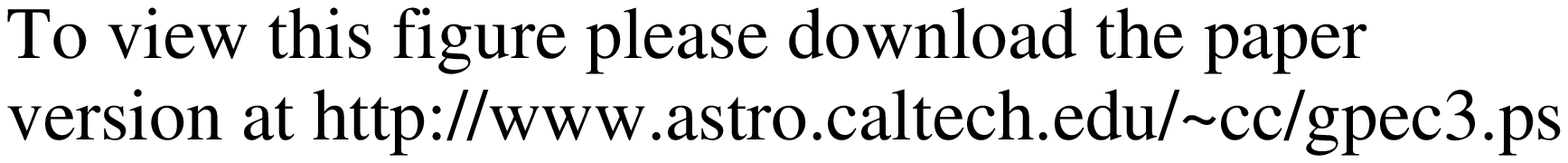}{6.0in}{0}{80}{80}{-250}{-100}
\vskip 0.1in
\caption{Montage images of the Perseus galaxies used in this 
study.   The number at the
upper left of each image is M$_{{\rm B}}$, while the lower number is the 
$(B-R)_{0}$ color, and the number to the far right is the catalog number from 
Table 1.  The solid black bar on each figure is 3\arcsec\, in length. {\bf
Please see http://www.astro.caltech.edu/$\sim$cc/gpec3.ps for this figure.}}
\end{figure}

\setcounter{figure}{0}
\clearpage
\begin{figure} 
\plotfiddle{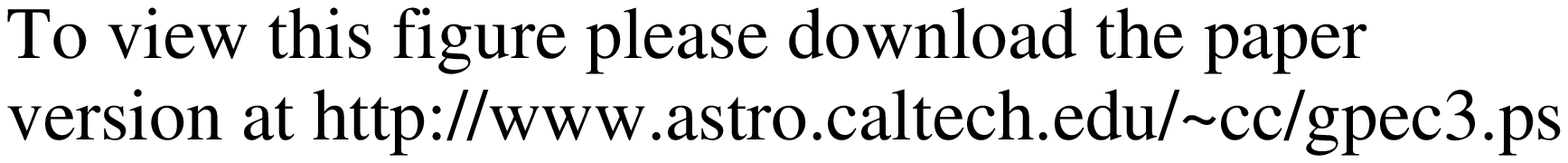}{6.0in}{0}{80}{80}{-250}{-100}
\vskip 0.1in
\caption{continued.}
\end{figure}

\setcounter{figure}{0}
\clearpage
\begin{figure} 
\plotfiddle{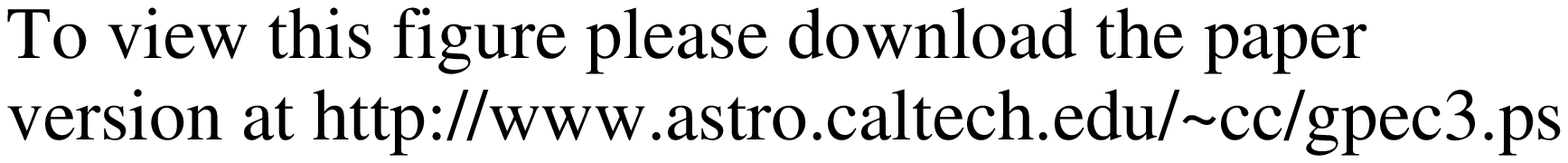}{6.0in}{0}{80}{80}{-250}{-100}
\vskip 0.1in
\caption{continued.}
\end{figure}

\setcounter{figure}{0}
\clearpage
\begin{figure}
\plotfiddle{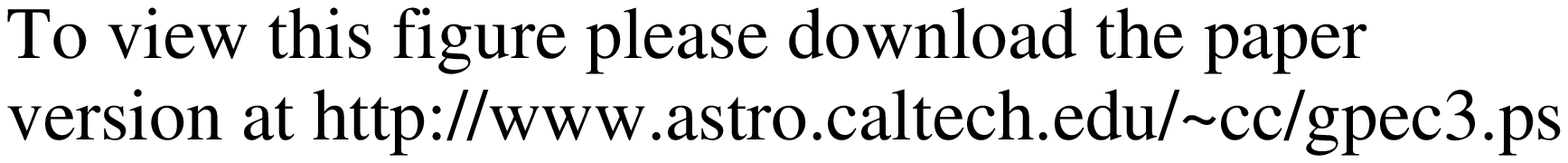}{6.0in}{0}{80}{80}{-250}{50}
\vskip -1in
\caption{continued.}
\end{figure}

\clearpage
\begin{figure}
\plotfiddle{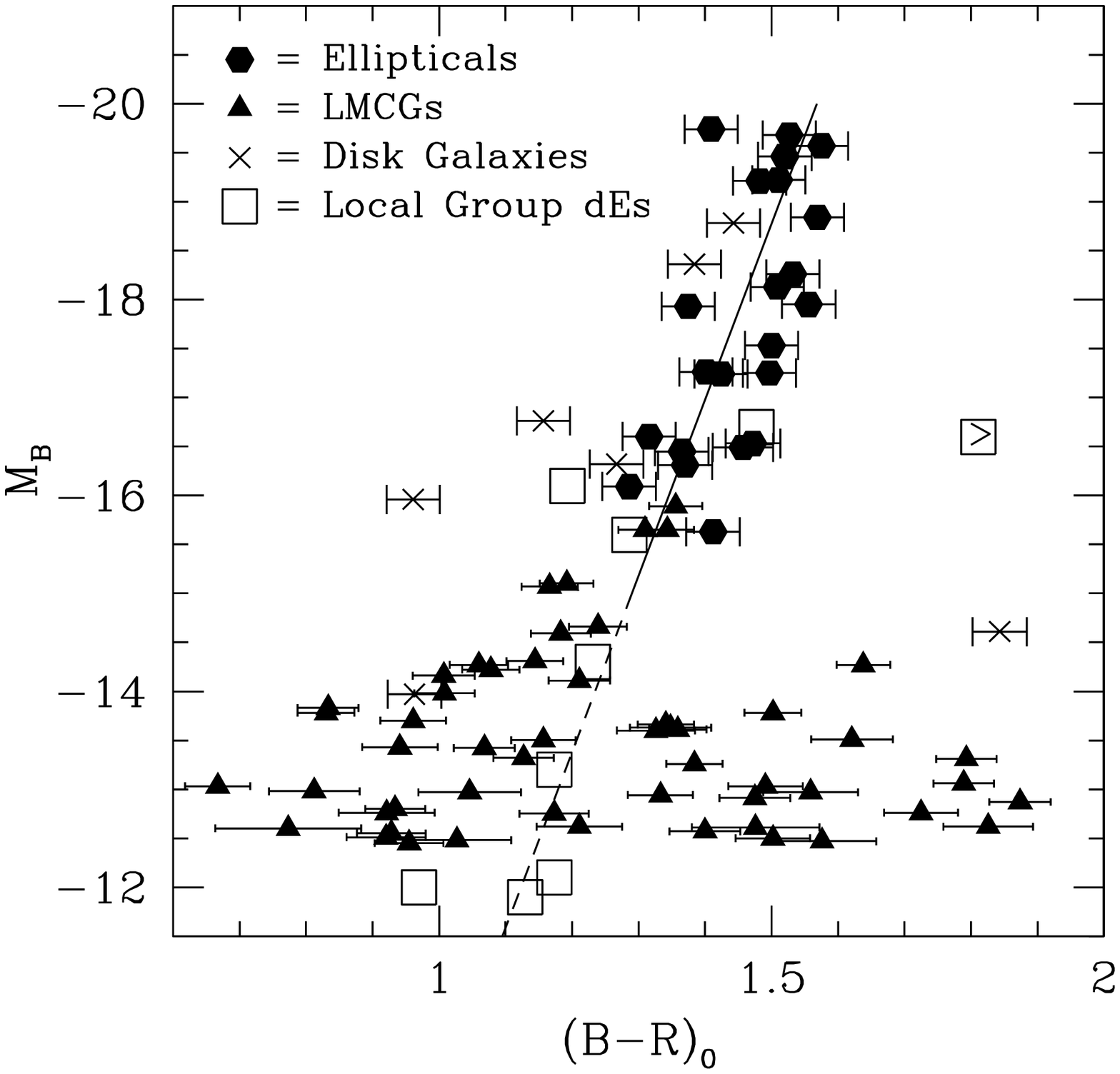}{6.0in}{0}{60}{60}{-190}{0}
\vskip -1in
\caption{The color magnitude diagram for galaxies in the Perseus 
cluster. Included here are objects with M$_{{\rm B}} <$ -12.5.  All 
the LMCGs in Figure~1 are plotted on this diagram.  The solid
line is the fit between color and magnitude for the brighter
objects with M$_{\rm B} < -16$ and the dashed line is the extension of
this fit to fainter magnitudes.   Also plotted as open boxes
are $(B-R)_{0}$ colors of Local Group dwarf ellipticals derived
from their metallicities using data from van den Bergh (2000). The LG dwarf
elliptical at $(B-R)_{0}\sim$1.8 and M$_{\rm B} \sim -16.5$ is an an upper 
limit. }
\end{figure}

\clearpage
\begin{figure}
\plotfiddle{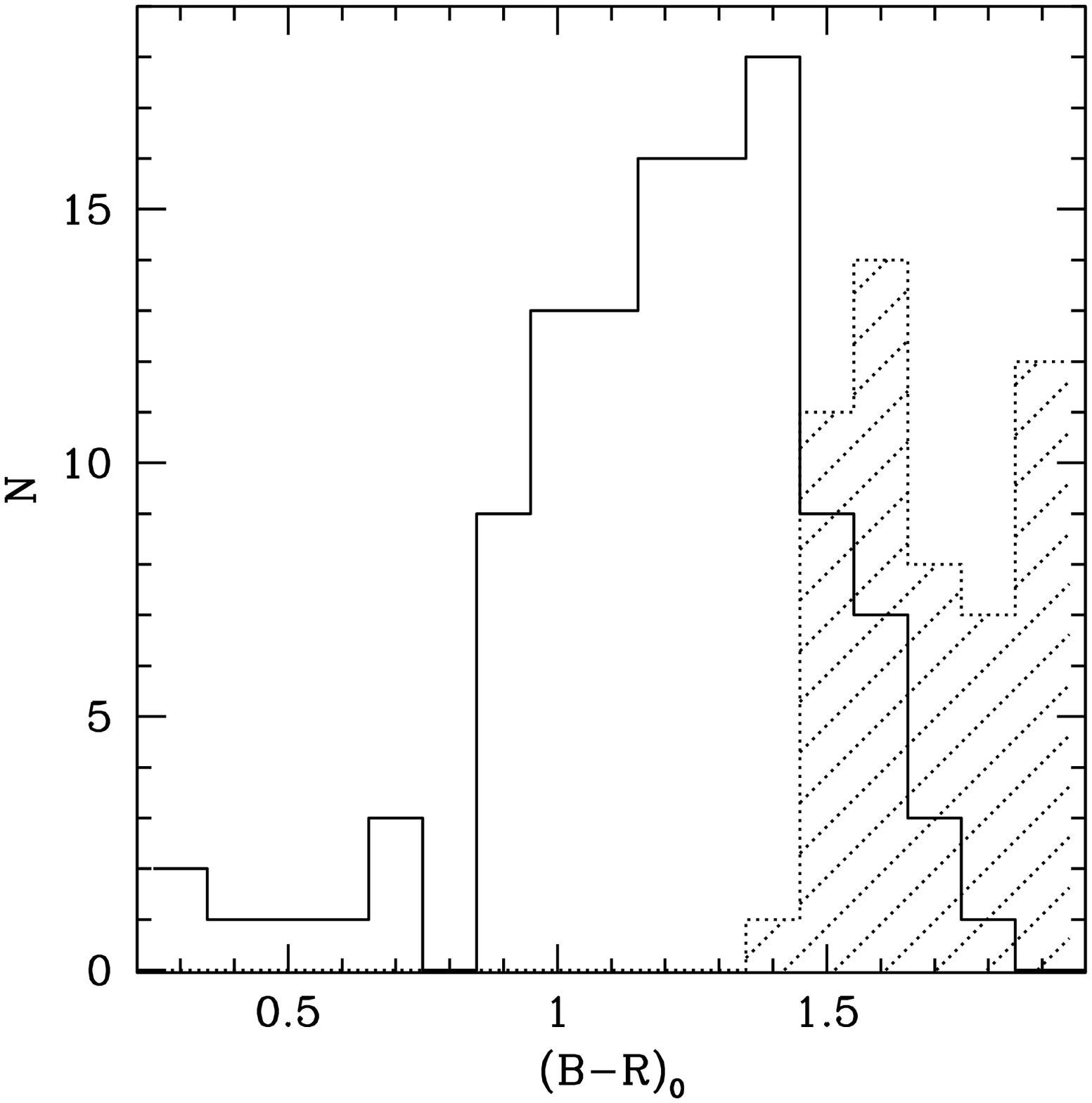}{6.0in}{0}{60}{60}{-190}{0}
\vskip -1in
\caption{Color distribution for all LMCGs in the Perseus cluster center
as identified in Paper II, down to M$_{\rm B} = -11$. The blue 
(solid) and red (shaded) LMCGs are plotted separately. }
\end{figure}

\clearpage
\begin{figure}
\plotfiddle{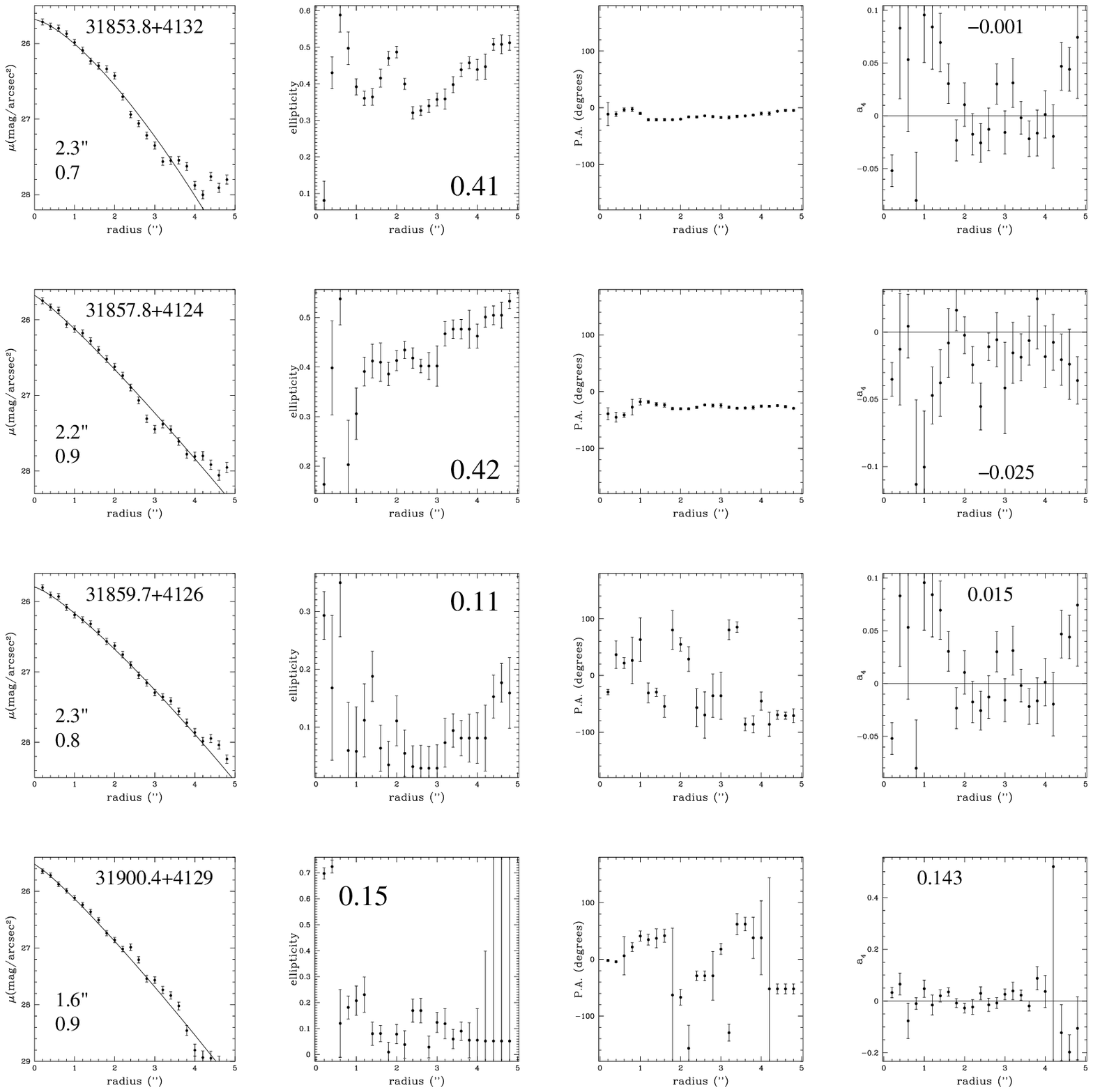}{7.0in}{0}{90}{90}{-280}{-30}
\vskip -1in
\caption{Surface brightness profiles, ellipticity profiles, 
position angles profiles and the a$_{4}$ Fourier component profiles 
for Perseus LMCGs
with M$_{\rm B} < -13.5$. The first panel displays the name, S\'{e}rsic profile
parameters, r$_{0}$ and n, for each galaxy.  The number in the second and
fourth column is the average ellipticities to 5\arcsec\, and 
a$_{4}$ value across all radii.}
\end{figure}

\setcounter{figure}{3}
\clearpage
\begin{figure}
\plotfiddle{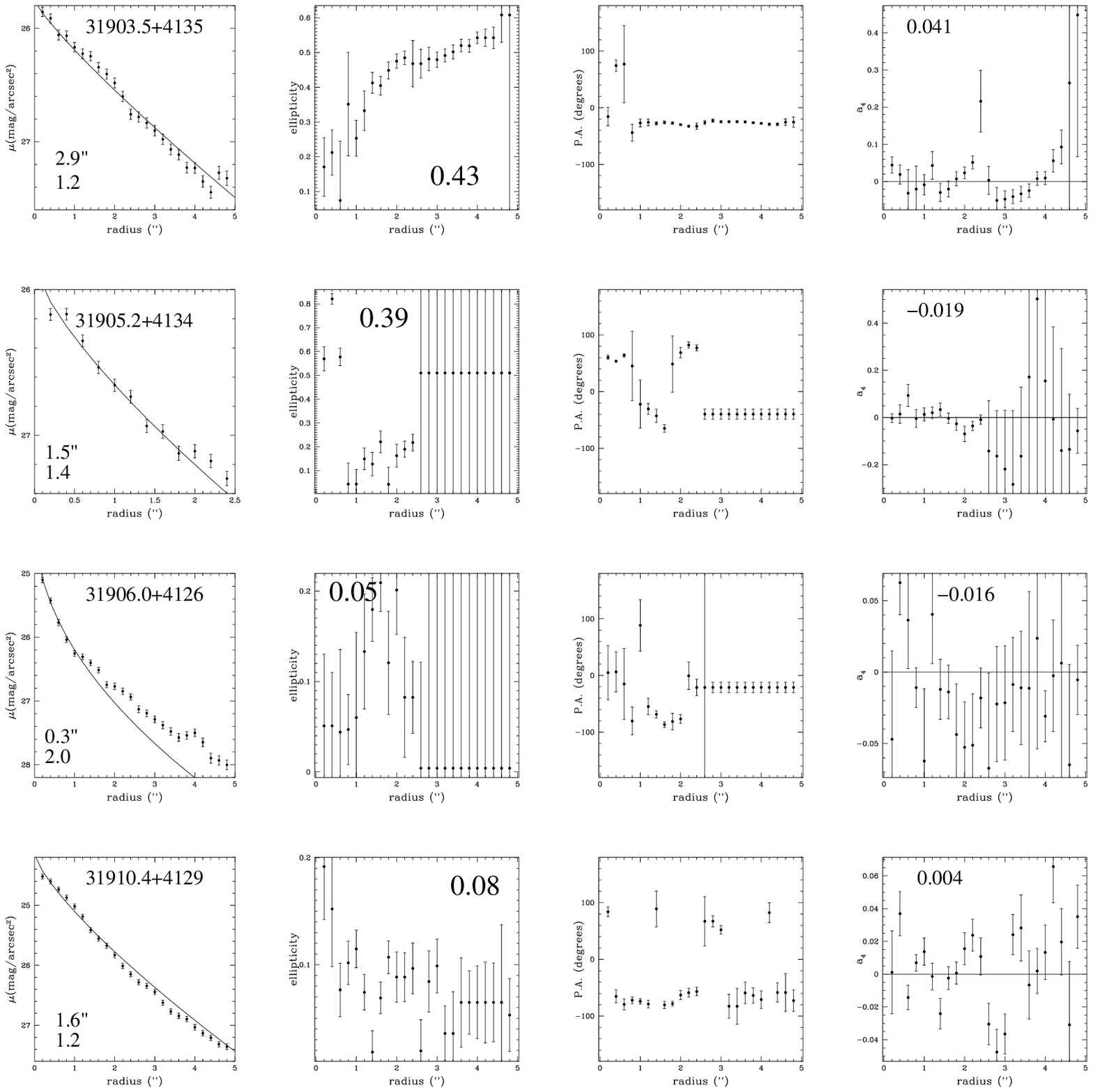}{7.0in}{0}{90}{90}{-280}{-30}
\vskip -1.5in
\caption{continued.}
\end{figure}

\setcounter{figure}{3}
\clearpage
\begin{figure}
\plotfiddle{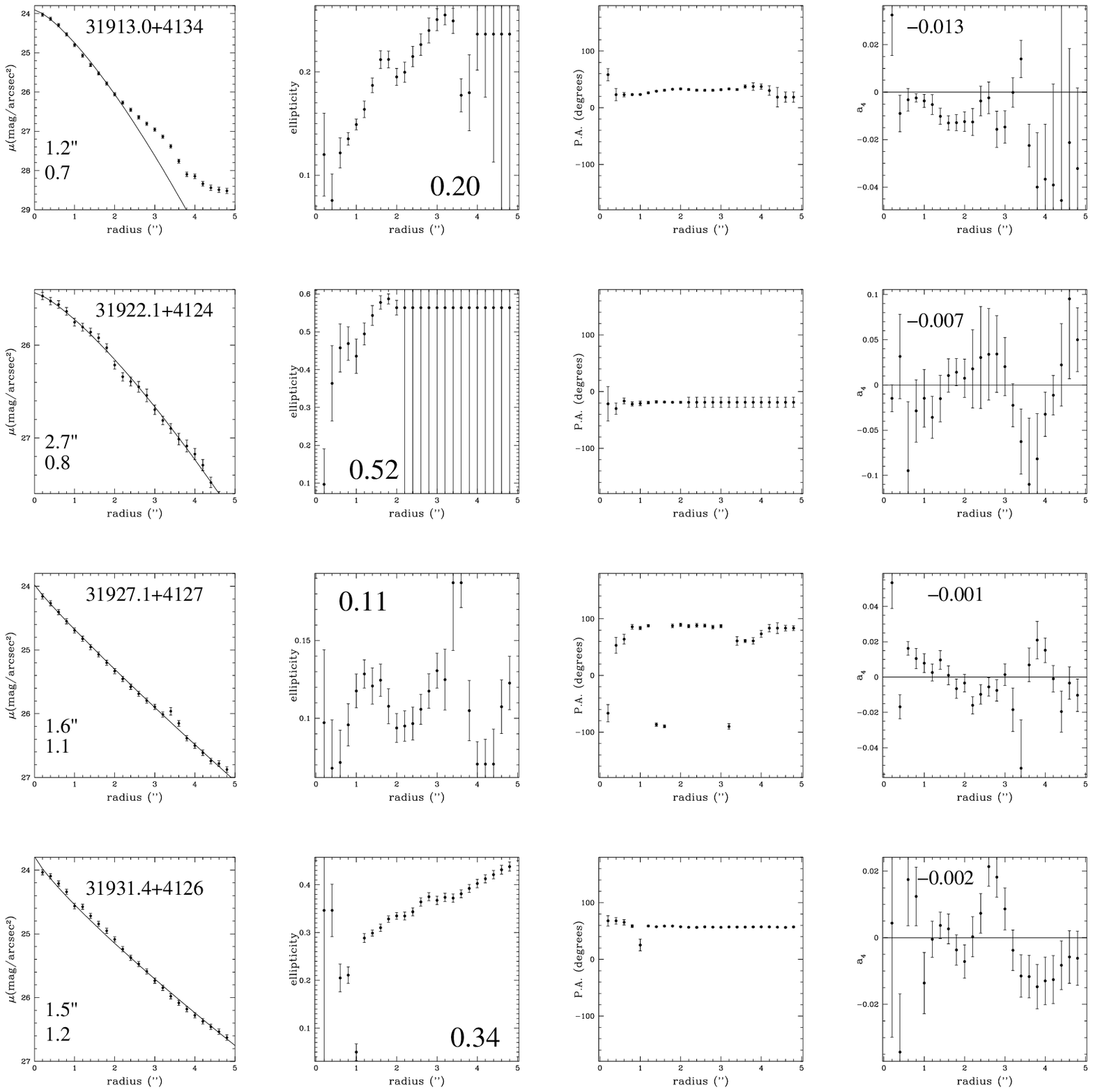}{7.0in}{0}{90}{90}{-280}{-30}
\vskip -1.5in
\caption{continued.}
\end{figure}
\clearpage

\setcounter{figure}{3}
\begin{figure}
\plotfiddle{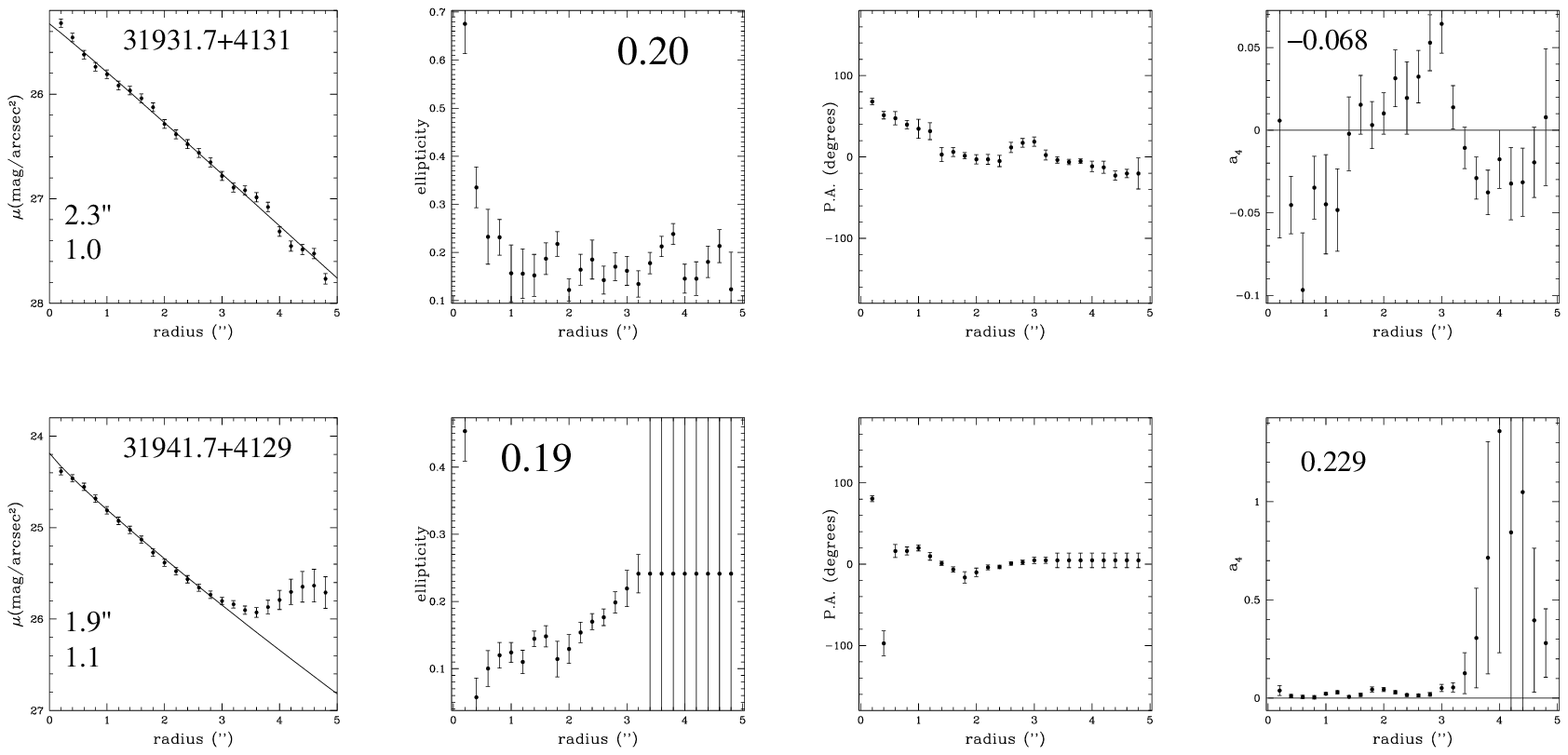}{6.0in}{0}{90}{90}{-280}{0}
\vskip -3.4in
\caption{continued.}
\end{figure}
\clearpage

\clearpage

\begin{figure}
\plotfiddle{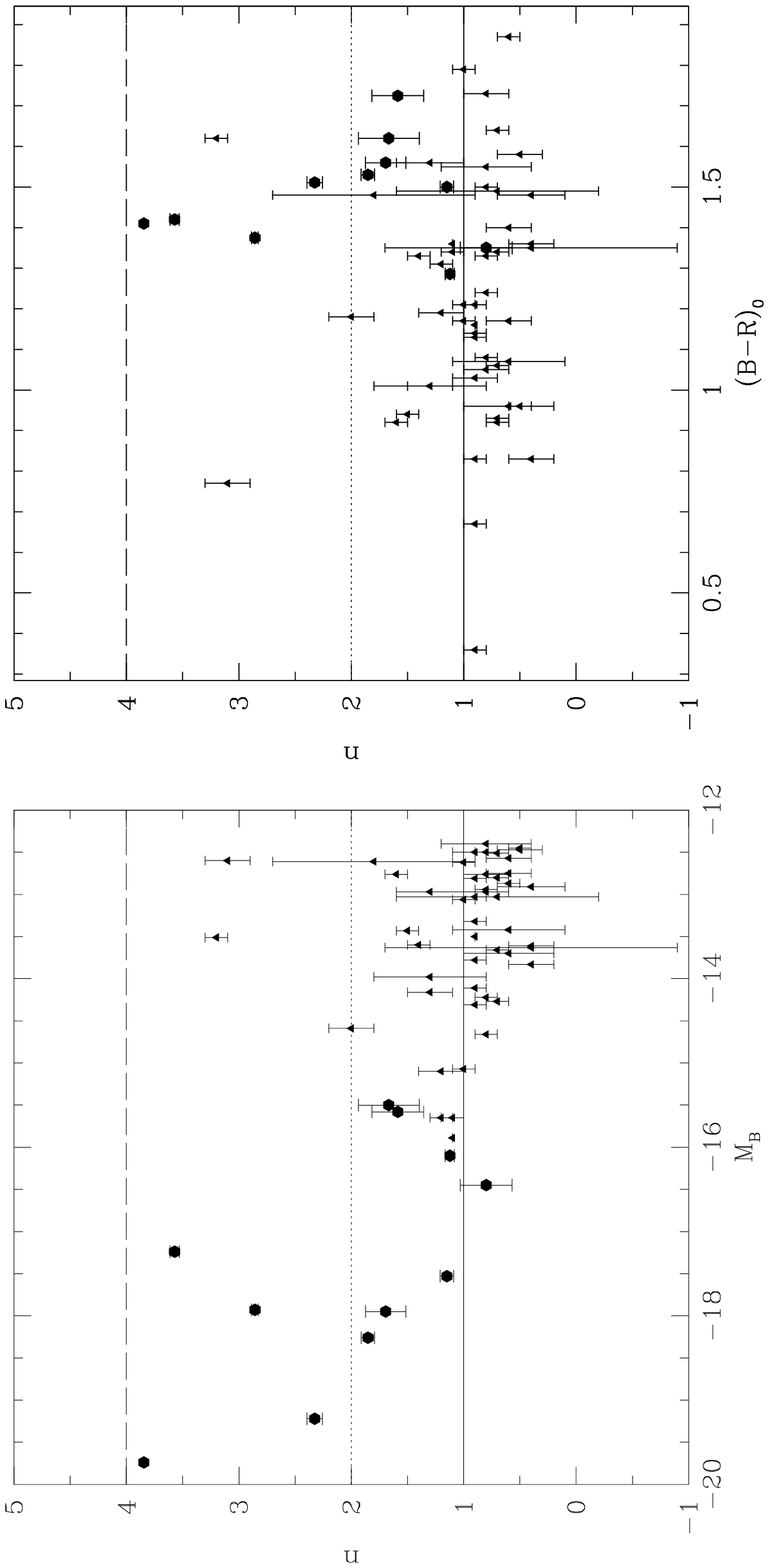}{6.0in}{0}{60}{60}{-190}{0}
\vskip 0in
\caption{S\'{e}rsic index n as a function of absolute
magnitude M$_{\rm B}$ (left panel) and $(B-R)_{0}$ color (right panel) 
for Perseus LMCGs.  
Symbols are the same as in Figure~2
with the triangle representing the LMCGs and the sexagons the giant
ellipticals.  The long dashed line shows the de Vaucouleur n = 4 limit
while the n = 2 limit for dwarf ellipticals is shown as the dashed line
and the the exponential profile of n = 1 is shown as the solid line.}
\end{figure}
\clearpage

\begin{figure}
\plotfiddle{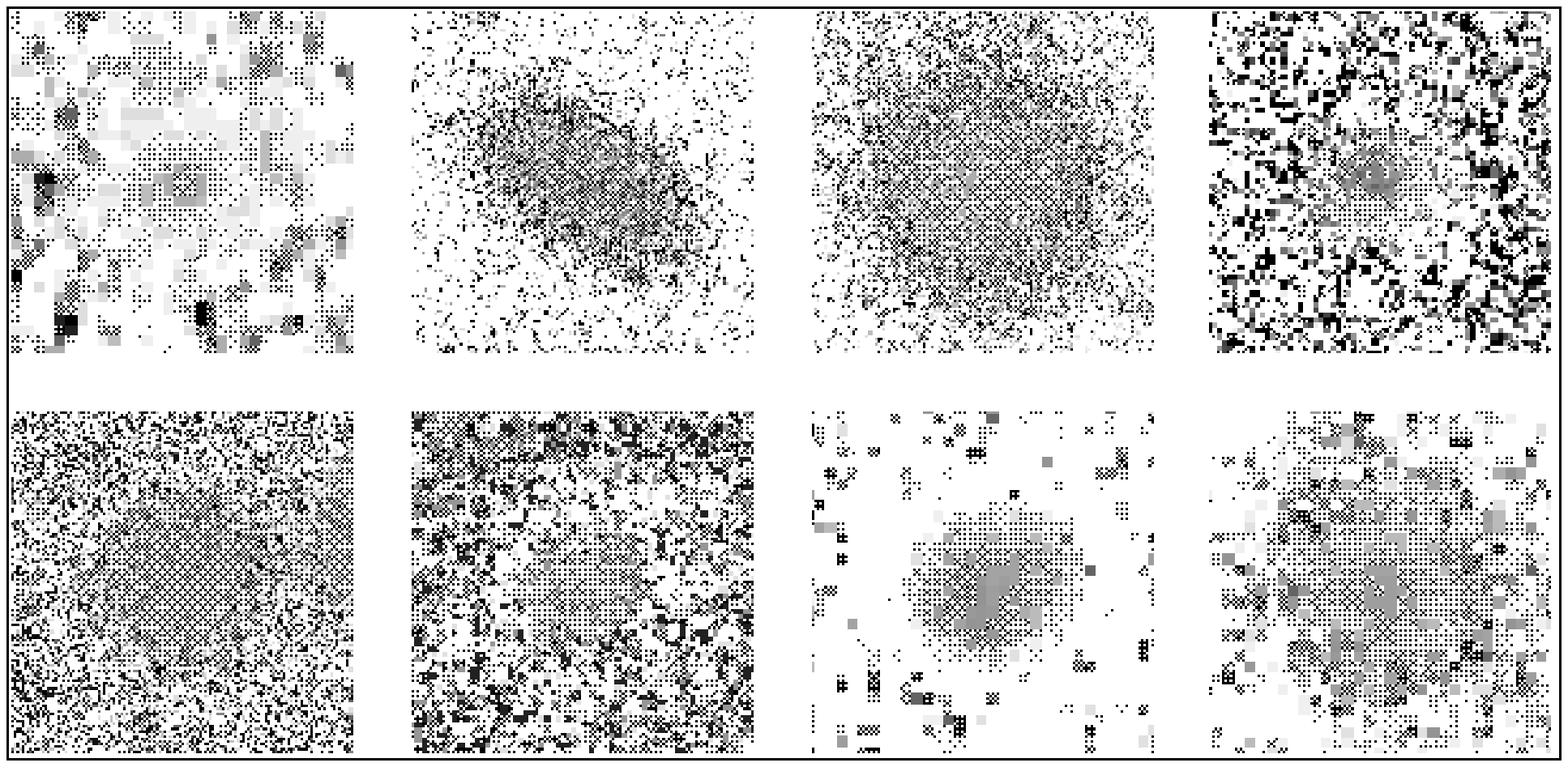}{7.0in}{0}{80}{80}{-250}{160}
\vskip -4.3in
\caption{$(B-R)$ color maps for bright Perseus LMCGs.  Several of 
these objects are nucleated, but none show large color gradients or distinct
core colors.  {\bf
Please see http://www.astro.caltech.edu/$\sim$cc/gpec3.ps for a much better
version of this figure.}}
\end{figure}

\clearpage

\begin{figure}
\plotfiddle{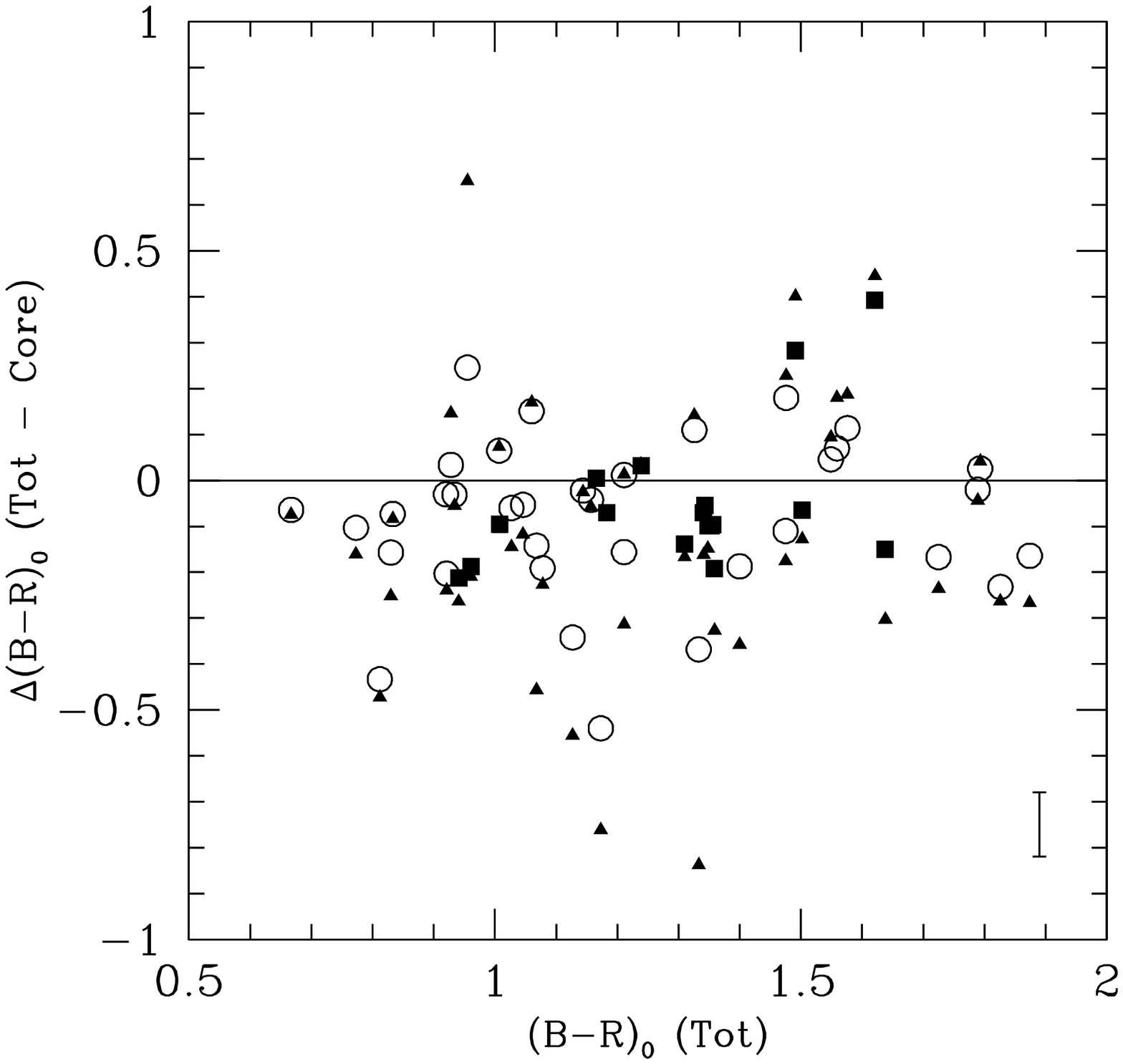}{6.0in}{0}{60}{60}{-190}{0}
\vskip -1.2in
\caption{The difference in $(B-R)_{0}$ color between LMCGs and 
their cores.  The solid boxes are nucleated LMCGs while the
open circles are the non-nucleated dEs. The solid
triangles represent the difference in $(B-R)_{0}$ between the outer region
of each LMCG and their core region.}
\end{figure}
\clearpage

\begin{figure}
\plotfiddle{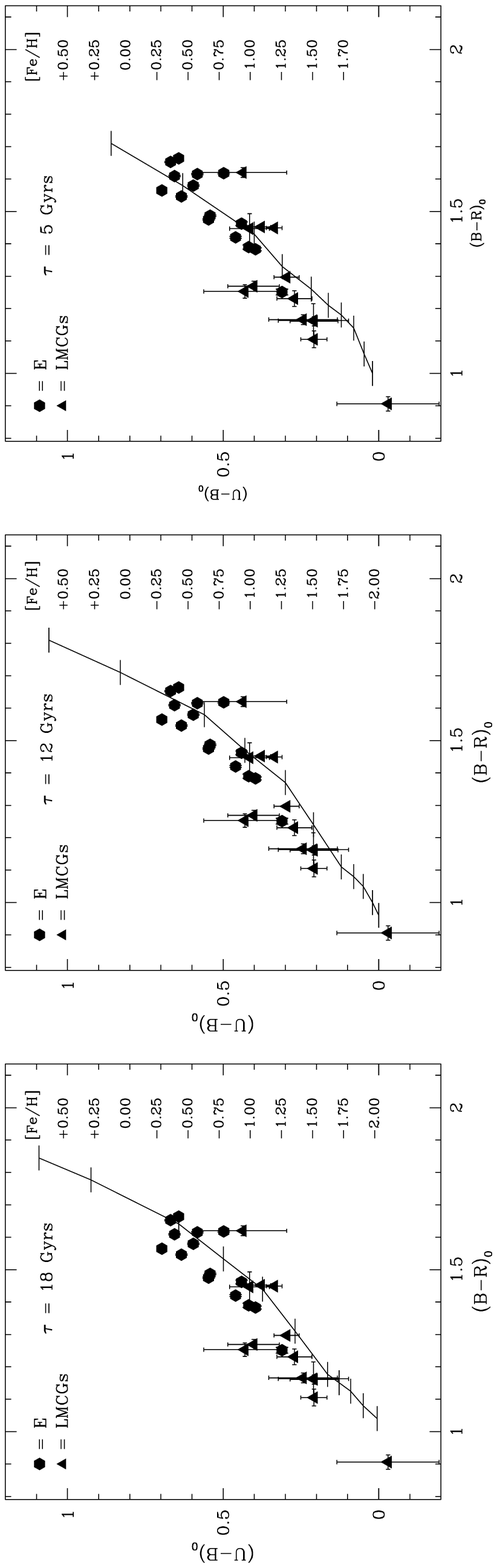}{6.0in}{0}{60}{60}{-190}{0}
\vskip 0in
\caption{UBR color-color diagram for Perseus cluster galaxies plotted
with Worthey isochrones at three different ages: 18~Gyr, 12~Gyr and 5~Gyr 
at metallicities from [Fe/H] = 0.5 to -2.  The horizontal lines along each 
isochrone
tick off the metallicities listed on the right hand side.}
\end{figure}

\begin{figure}
\plotfiddle{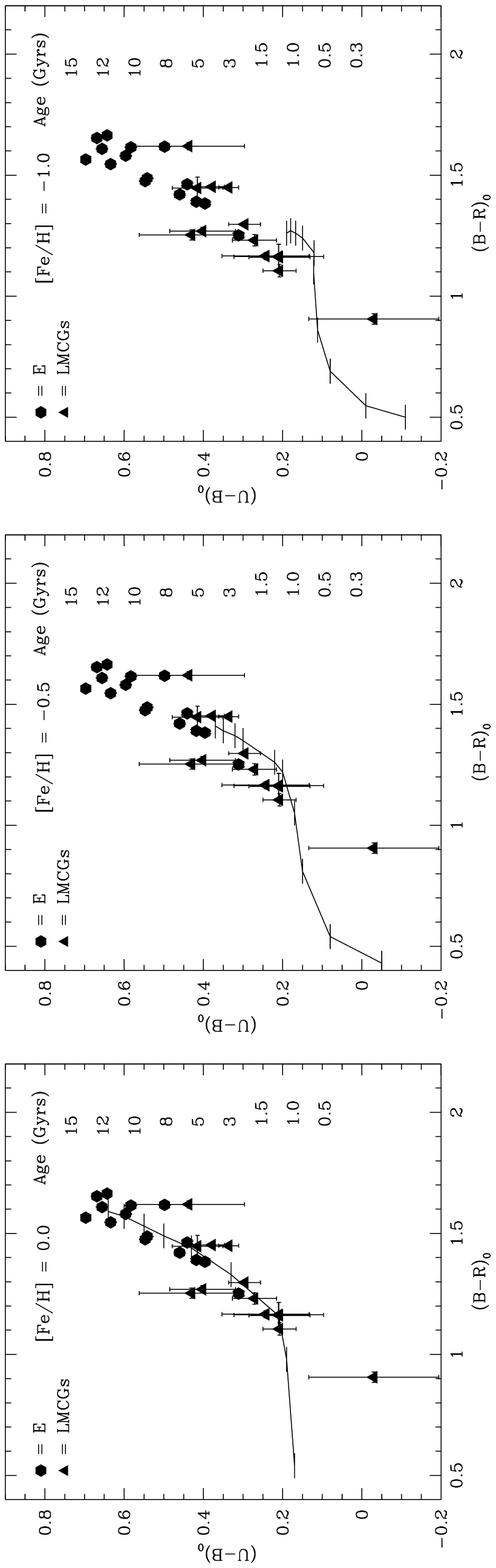}{6.0in}{0}{60}{60}{-190}{0}
\vskip 0in
\caption{Three stellar synthesis modeled age tracks on the UBR diagram at
constant metallicities of solar, [Fe/H] = -0.5 and [Fe/H] = -1. The age
range is 0.3 Gyrs to 15 Gyrs for the [Fe/H] = -0.5 and -1
models and 0.5 Gyrs to 15 Gyrs for the solar metallicity models.}
\end{figure}

\begin{figure}
\plotfiddle{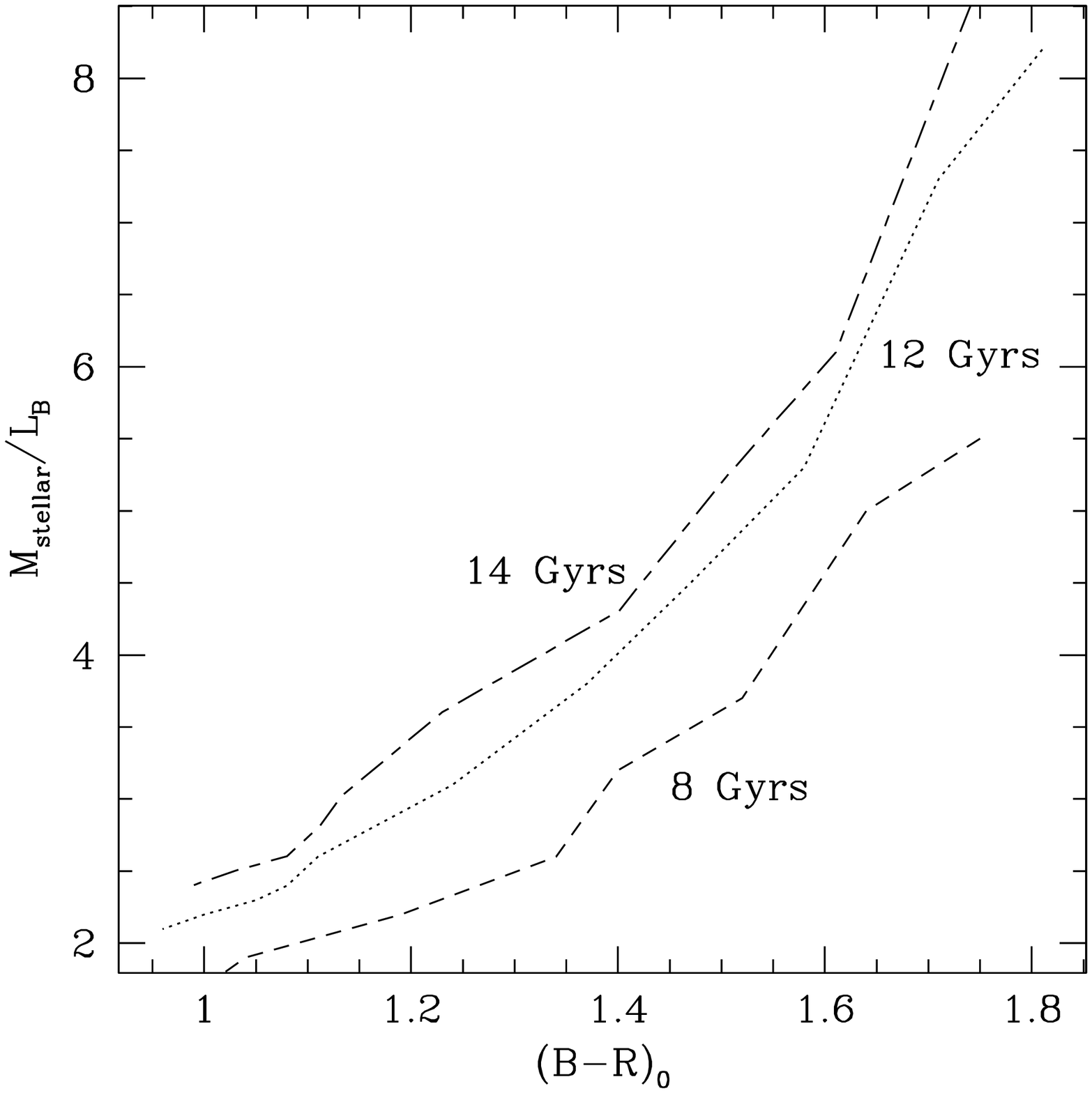}{6.0in}{0}{60}{60}{-190}{0}
\vskip -1.2in
\caption{The relationship between $(B-R)_{0}$ color and $M_{\rm stellar}/L_{B}$
ratios at three different ages using the stellar synthesis models of
Worthey (1994) and assuming that color is a measure of metallicity.}
\end{figure}

\begin{figure}
\plotfiddle{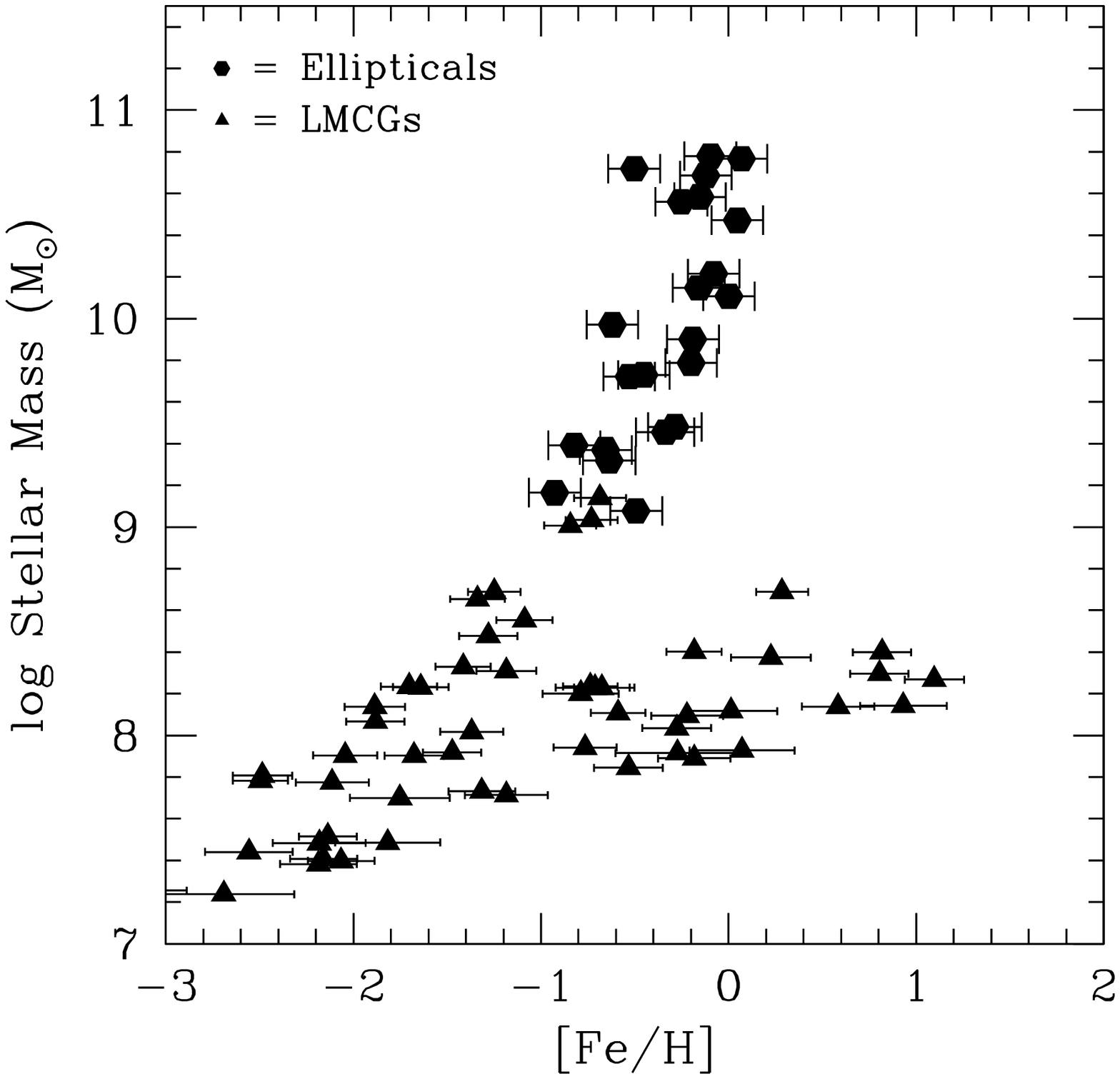}{6.0in}{0}{60}{60}{-190}{0}
\vskip -1.2in
\caption{The relationship between stellar mass and metallicity for
ellipticals and LMCGs in the Perseus cluster.  These properties are
found by converting $(B-R)_{0}$ color into metallicity and luminosity
into stellar mass by assuming a M$_{\rm stellar}$/L given by the color
(see Text).}
\end{figure}

\begin{figure}
\plotfiddle{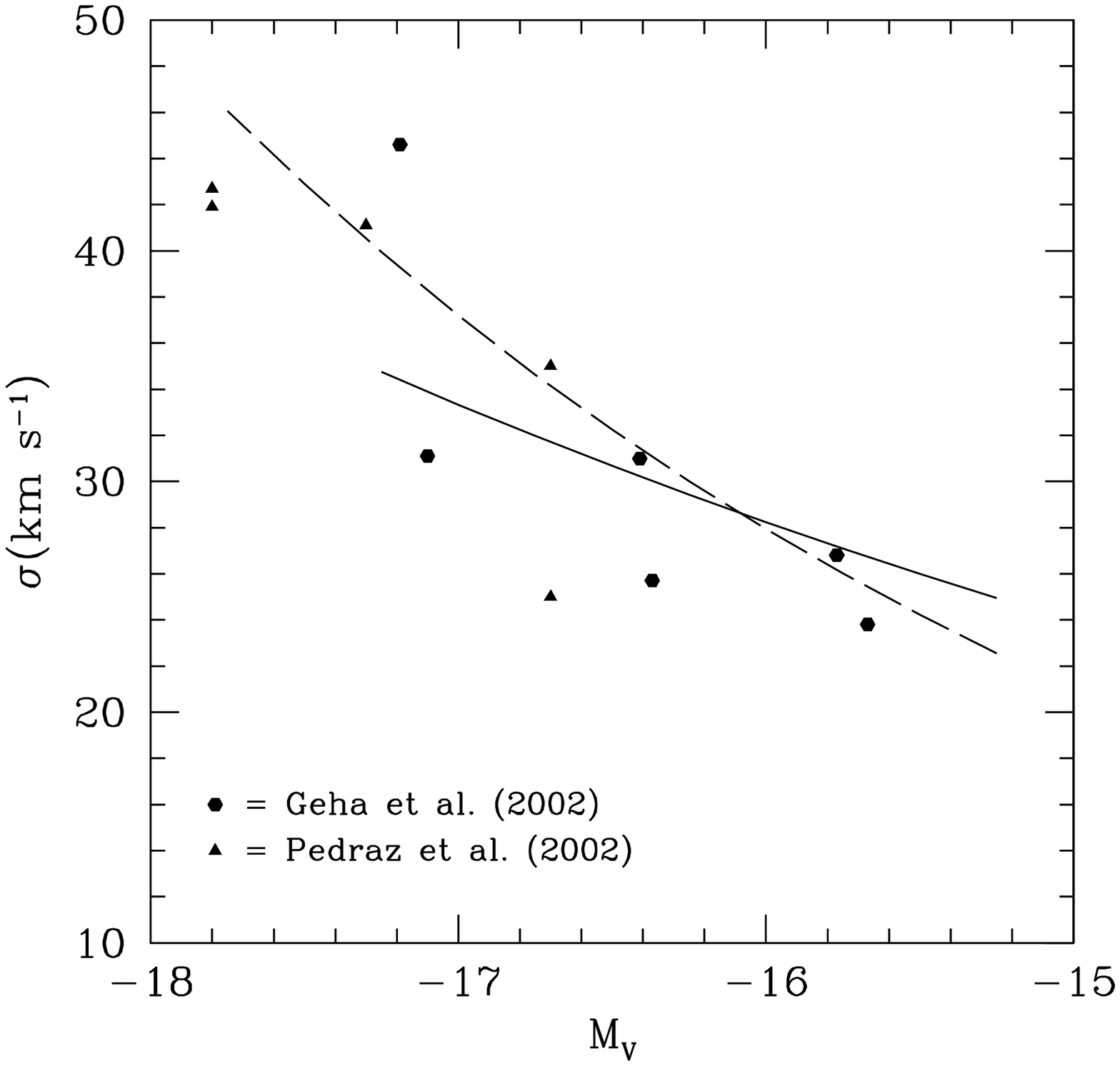}{6.0in}{0}{60}{60}{-190}{0}
\vskip -1.2in
\caption{The relationship between absolute magnitude M$_{\rm V}$ and
central velocity dispersion $\sigma$ for Virgo cluster galaxies
observed by Geha et al. (2002) and Pedraz et al. (2002).  The solid line
shows the fit between these two parameters for systems at M$_{\rm V} > -17$
while the dashed line is the fit for all the galaxies.}
\end{figure}

\begin{figure}
\plotfiddle{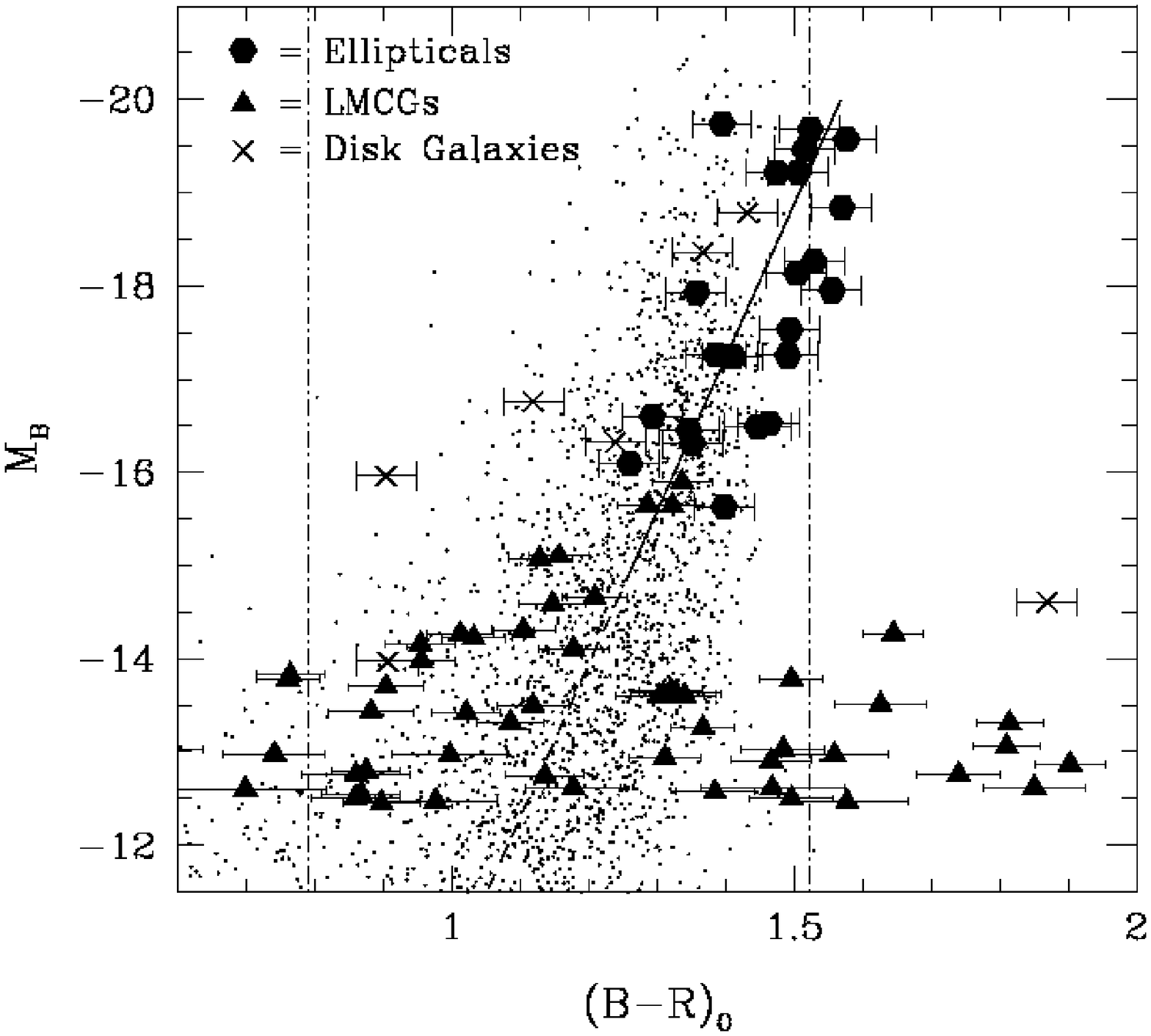}{6.0in}{0}{60}{60}{-190}{0}
\vskip -1in
\caption{The color-magnitude diagram of Perseus galaxies plotted
with simulated $\Lambda$CDM galaxies from Nagamine et al. (2001).
The vertical dot-dashed line shows the limitations of the $(B-R)_{0}$ 
metallicity calibration from globular clusters. }
\end{figure}

\clearpage

\begin{figure}
\plotfiddle{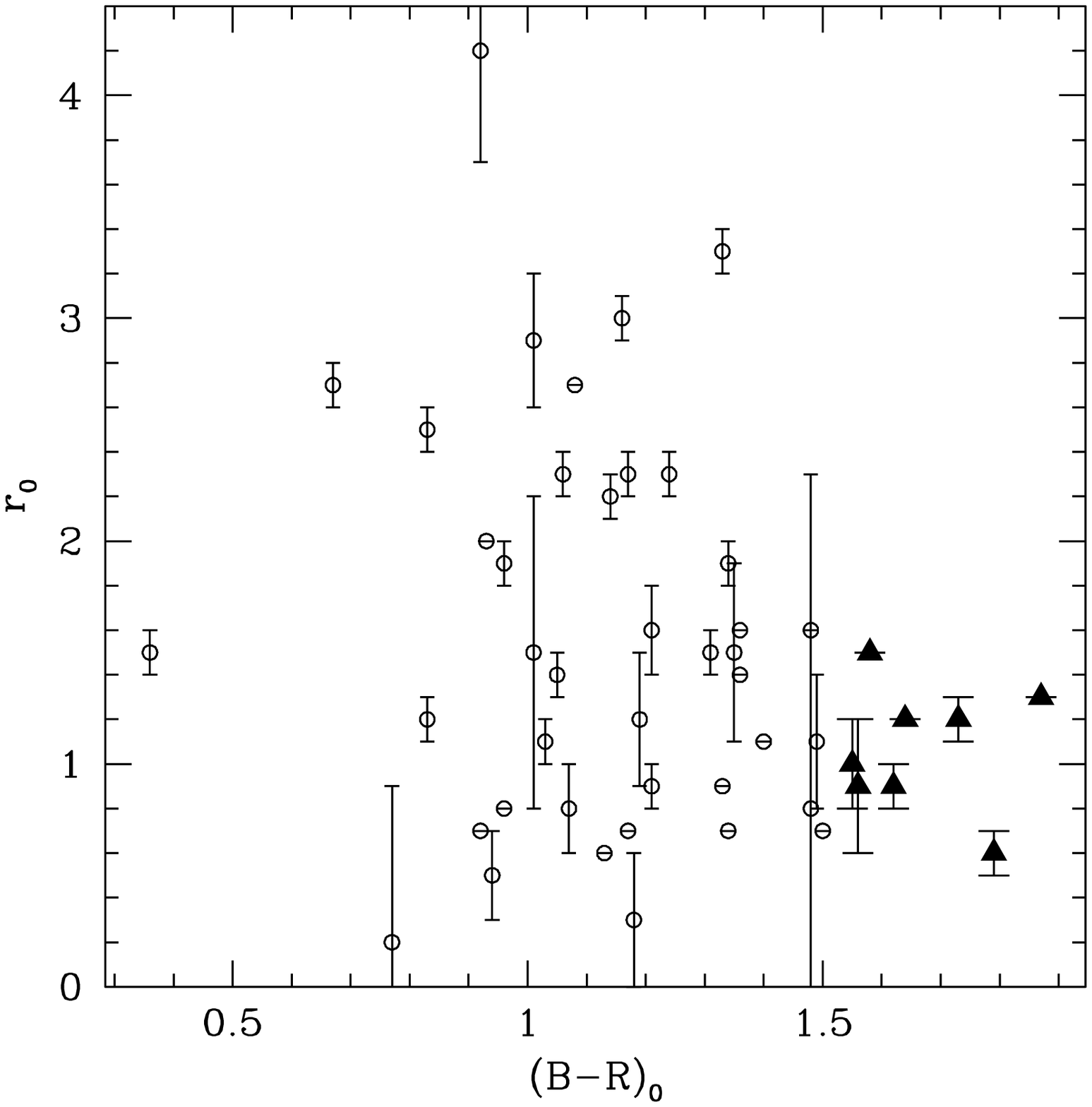}{6.0in}{0}{60}{60}{-190}{0}
\vskip -1in
\caption{Relationship between the S\'{e}rsic scale length r$_{0}$ and
$(B-R)_{0}$ color for Perseus LMCGs.  The open circles are for the blue 
LMCGs while the triangles are the red LMCGs.}
\end{figure}

\begin{figure}
\plotfiddle{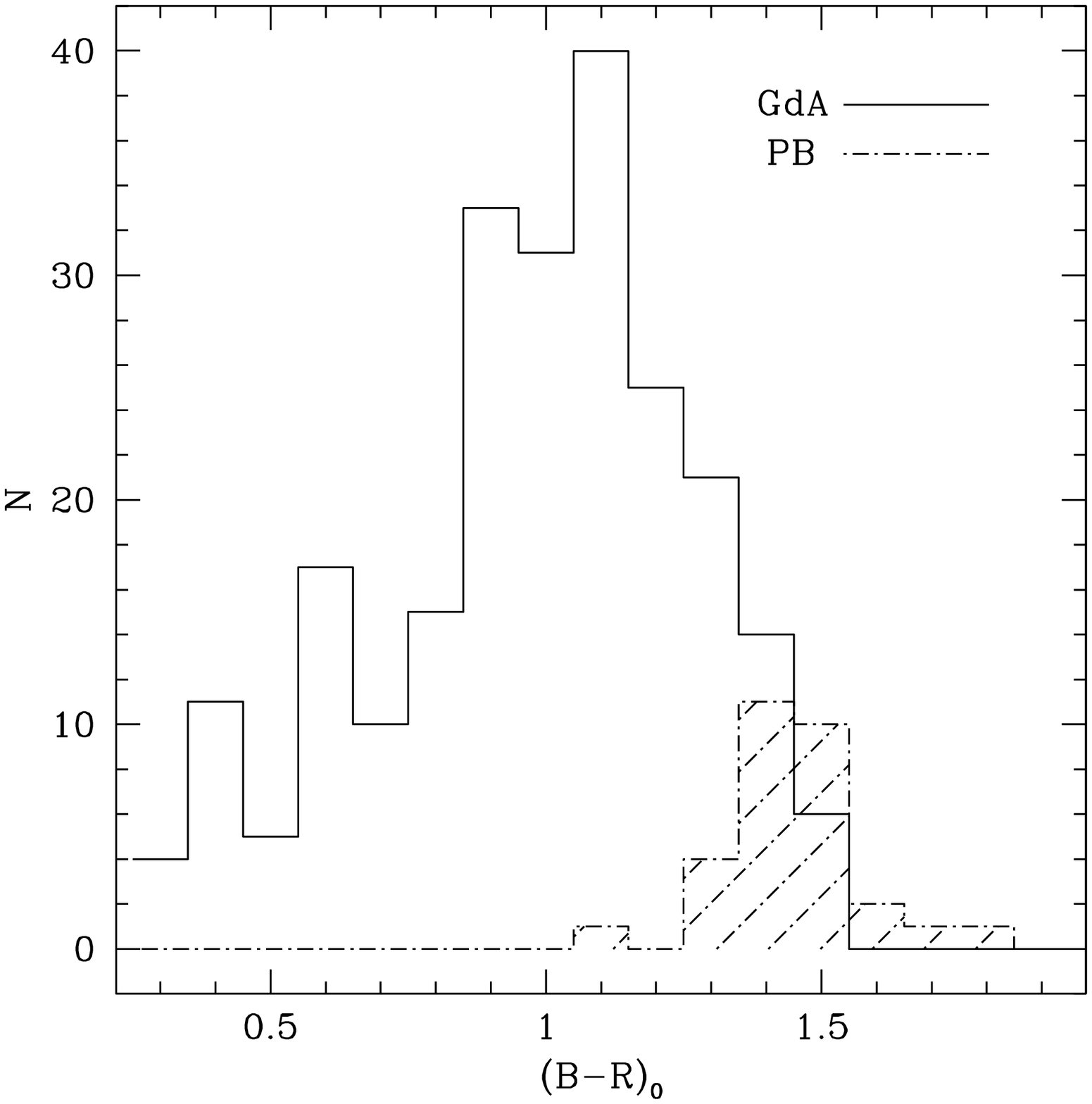}{6.0in}{0}{60}{60}{-190}{0}
\vskip -1in
\caption{Histogram of $(B-R)_{0}$ colors for spiral galaxy bulges taken
from the data in Gadotti \& dos Anjos (2001) (GdA) and Peletier \&
Balcells (1997) (PB).}
\end{figure}

\end{document}